\documentclass[prb,floats,twocolumn,floatfix,superscriptaddress]{revtex4}
\usepackage{graphicx}
\usepackage{color}
\usepackage{ulem}
\newcommand{\bq}{{\bf q}}
\newcommand{\bk}{{\bf k}}
\newcommand{\br}{{\bf r}}
\newcommand{\reff}[1]{{(\ref{#1})}}
\newcommand{\chibf}{\chi\hspace{-2mm}\chi}
\begin{document}
\title{Spatially Modulated Electronic Nematicity in the Three-Band Model of 
Cuprate Superconductors}
\date{\today}
\author{S. Bulut} 
\affiliation{Department of Physics and Astronomy, Trent
  University, Peterborough Ontario, Canada, K9J 7B8} 
\affiliation{Department of Physics, Queen's University, 
Kingston Ontario, Canada, K7L 3N6}
\author{W. A. Atkinson} 
\email{billatkinson@trentu.ca}
\affiliation{Department of Physics and Astronomy, Trent
  University, Peterborough Ontario, Canada, K9J 7B8} 
\author{A. P. Kampf\,} \affiliation{Theoretical Physics III,
Center for Electronic Correlations and Magnetism, Institute of Physics,
University of Augsburg, 86135 Augsburg, Germany}
\begin{abstract}
Charge order in cuprate superconductors is a possible source of anomalous 
electronic properties in the underdoped regime. Intra-unit cell charge ordering
tendencies point to electronic nematic order involving oxygen
orbitals.  In this context we investigate charge instabilities in the
Emery model and calculate the charge susceptibility within diagrammatic 
perturbation theory. In this approach, the onset of charge order is signalled 
by a divergence of the susceptibility. Our calculations reveal three different 
kinds of order: a commensurate ($q=0$) nematic order, and two incommensurate 
nematic phases with modulation wavevectors that are either axial or oriented 
along the Brillouin zone diagonal.  We examine the nematic phase diagram as a 
function of the filling, the interaction parameters, and the band structure. We 
also present results for the excitation spectrum near the nematic instability, 
and show that a soft nematic mode emerges from the particle-hole continuum at 
the transition. The Fermi surface reconstructions that accompany the modulated 
nematic phases are discussed with respect to their relevance for 
magneto-oscillation and photoemission measurements. The modulated nematic 
phases that emerge from the three-band Emery model are compared to those found 
previously in one-band models.
\end{abstract}
\maketitle
\section{Introduction}

Cuprate superconductors are susceptible to spin and charge density ordered 
phases that compete with superconductivity. This is well established in 
La-based cuprates,\cite{Tranquada:1995,Abbamonte:2005,Vojta:2009} where (quasi)
static spin/charge stripes are widely observed, even in coexistence with 
superconductivity\cite{Tranquada:2008,Hucker:2012tv}. However, 
their presence in other cuprate families is generally unconfirmed.
Because density waves are one of the proposed origins for the pseudogap in the 
underdoped regime, it is necessary to establish whether charge and spin order
are universal amongst the cuprates.  The recent discovery of charge
order (without spin order) in underdoped $\mathrm{YBa_2Cu_3O_{6+x}}$
(YBCO) is an important step in this direction.  In this work, we
describe novel incommensurate charge-ordered phases that arise in the
three-band Emery model for cuprates, and discuss the extent to which these
are consistent with the charge order detected in YBCO.

Early evidence for broken symmetry phases in YBCO came from magneto-oscillation 
experiments, which identified a reconstructed Fermi
surface\cite{DoironLeyraud:2007bj,Sebastian:2012wh} with an electron-like Fermi 
surface pocket\cite{LeBoeuf:2007gi} that emerges when strong magnetic fields 
are applied. These experiments were theoretically discussed from the perspective
of density waves.\cite{Chakravarty:2011} Nernst effect measurements found a 
uniaxial symmetry breaking,\cite{Daou:2010bo} consistent with a charge-density 
wave (CDW). Subsequent nuclear magnetic resonance (NMR) experiments further 
suggested a commensurate CDW with a period of four unit cells, with no evidence 
for any spin ordering\cite{Wu:2011ke}.  In this work, the authors made a clear 
distinction between ortho-II YBCO (with hole doping $p\sim 0.108$--0.12) where 
{\it only} charge order is seen, and lower dopings near the 
superconductor-insulator transition where charge and spin order are both 
seen.\cite{Hinkov:2007gq,Hinkov:2008sci,Haug:2010ei}
 
More recently, x-ray
scattering\cite{Ghiringhelli:2012bw,Chang:2012vf,Blackburn:2012wn_prl}
experiments have identified a CDW phase in the same doping and
magnetic field range in which Fermi surface pockets are detected. The
charge pattern is aligned with the crystalline
axes\cite{Ghiringhelli:2012bw,Chang:2012vf} and is incommensurate,
with a weakly doping dependent period of $\sim 3.2$ lattice constants.
Whether this CDW is uniaxial\cite{Daou:2010bo,Wu:2011ke} or
biaxial\cite{LeBoeuf:2012up_nphys,Blanco-canosa:2013} has not been resolved, and may depend on
doping.\cite{Blackburn:2012wn_prl}
Regardless of the details, X-ray and NMR experiments  established that the incommensurate CDW competes with the superconductivity,\cite{Chang:2012vf,Wu:2013} implying that both phases operate on similar energy scales.

The x-ray results suggest that this charge-ordered state is distinct from the 
stripe phase in La-based cuprates;\cite{Blackburn:2012wn_prl,Hucker:2012tv,Blanco-canosa:2013} 
however, its relation to apparent charge-ordered phases in Bi-based cuprates 
is still not clear. Photoemission experiments on underdoped Bi-cuprates have 
found spectral features\cite{Kondo:2009,Hashimoto:2010nphys,He:2011,Ideta:2012} 
that are consistent with competing non-superconducting phases. Scanning 
tunneling microscopy (STM) experiments\cite{Hoffman:2002,Howald:2003,Wise:2008,Parker:2010nat,Lawler:2010n,Mesaros:2011s,Eduardo:2012prb}
further pointed to intra-unit cell charge order. The simplest candidates for 
such order are uniaxial ``nematic'', or biaxial ``checkerboard'' phases 
involving a spontaneous transfer of charge between oxygen $p_x$ and $p_y$ 
orbitals within the CuO$_2$ unit cell.
  
In the present work, we report on possible charge-ordered phases within the 
three-band Emery model,\cite{Emery:1987prl} which includes copper $d$ and oxygen
$p$ orbitals. We have calculated the charge susceptibility 
$\chi_{\alpha\beta}(\bq,\omega)$ ($\alpha$ and $\beta$ are orbital indices)  
diagrammatically.\cite{Littlewood:1989,Littlewood:1990} The leading 
instabilities of the model are obtained from divergences of the static 
susceptibility $\chi_{\alpha\beta}(\bq)$. We find that, through much of the 
phase diagram, the leading instability is to an incommensurate (finite-$q$) 
charge modulation involving primarily the oxygen orbitals. When the ordering 
wavevector ${\bf q}^*$ tends to zero, this phase evolves continuously into the 
commensurate nematic phase proposed by Fischer and Kim\cite{Fischer:2011} to 
explain the STM results for Bi-cuprates.

The possibility of finite-$q$ ``modulated nematic order'' (MNO) was raised 
previously in one-band 
models.\cite{Metlitski:2010vf,Melikyan:2011,Husemann:2012vg,Holder:2012ks,Efetov:2013,Sachdev:2013,Meier:2013} In this context, 
``nematic'' refers to an anisotropic renormalization of the bond-centered 
kinetic energy that disrupts the tetragonal symmetry of the lattice. This anisotropy
can be viewed as the result of integrating out inequivalent oxygen orbitals in a
three-band model,\cite{Efetov:2013} and in this light the one-band bond-centered nematicity
may be related to  the nematic phase reported here, which is
characterized by a spontaneous transfer of charge between O$p_x$ and O$p_y$ 
orbitals. Note that we make a distinction between MNO, for which the charge 
transfer is predominantly intra-unit cell, and conventional CDW order, for 
which the charge transfer is inter-unit cell.


Despite the differences between one- and three-band models, there are some 
surprising similarities in the structure of the modulations. Within the 
``hotspot'' model\cite{Metlitski:2010vf,Efetov:2013} the modulation
wavevector $\bq^\ast$  lies along the Brillouin zone diagonal. Holder and
Metzner\cite{Holder:2012ks} considered a more general interaction and found 
that $\bq^\ast$ may be either diagonal or axial (bond-aligned), depending on
the Fermi-surface shape. Here, we find that the doping dependence of $\bq^\ast$ 
is qualitatively the same as that found by Holder and Metzner, even though the 
mechanism which drives the instability is different.

Earlier, a different charge instability, involving charge transfer between Cu 
and O was found in the three-band Emery model when the energies of O$p_x$, 
O$p_y$, and Cu$d$ orbitals are close to 
degenerate.\cite{Littlewood:1989,Littlewood:1990,Scalettar:1991} This charge
transfer instability does not occur for the more realistic range of parameters 
studied below.  In the strong-coupling limit of the Emery model it was instead argued that 
the interaction $V_{pp}$ between neighboring oxygen sites confines the motion of doped holes to one-dimensional channels which 
thereby suggests a possible source for nematicity\cite{Kivelson:2004}. 
A continuum theory for the quantum phase transition to the nematic state was 
subsequently developed\cite{Sun:2008}.


We describe the model and the diagrammatic calculations in detail in 
Sec.~\ref{sec:model}; results are presented in Sec.~\ref{sec:results}. From the 
static nematic susceptibility, we obtain phase diagrams as functions of 
temperature, doping, and interaction strengths. We show that, for realistic 
model parameters, there exists a wide doping range over which incommensurate 
nematic order is preferred over commensurate order. The results for the 
dynamical susceptibility show that the nematic transition is marked by a soft 
mode that emerges from the particle-hole continuum near the nematic 
instability. We also describe the expected Fermi-surface reconstruction in the 
nematic phase. In Sec.~\ref{sec:discussion} we compare one- and three-band 
models showing why they generate similar phase diagrams despite significant 
differences in the models. Finally we discuss to what extent our results for 
modulated nematic order are compatible with the existing experimental evidence 
for charge ordering.

\section{Model and Calculations}
\label{sec:model}
In this section we describe briefly the three-band Emery 
model\cite{Emery:1987prl} and outline the calculation of the charge 
susceptibility matrix $\chi_{\alpha\beta}(\bq,\omega)$.  

\subsection{Hamiltonian}
\begin{figure}
 \includegraphics[width=0.6\columnwidth]{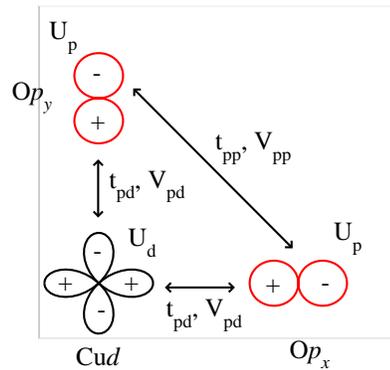}
 \caption{(Color online) Unit cell for the three-band Emery model. The model 
    includes one copper $d_{x^2-y^2}$ orbital along with two oxygen orbitals, 
    denoted by O$p_x$ and O$p_y$. Hopping matrix elements ($t_{pd}$ and
    $t_{pp}$) and Coulomb interaction strengths ($U_d$, $U_p$, $V_{pd}$,
    $V_{pp}$) are indicated.  Throughout we fix $t_{pd}=1$, $U_d=9$, $U_p=3$, 
    and $V_{pd}=1$.  }
  \label{fig:unitcell}
\end{figure}

The unit cell of a single CuO$_2$ plane is illustrated in 
Fig.~\ref{fig:unitcell}. It consists of three orbitals, one copper $d_{x^2-y^2}$ 
orbital and two oxygen orbitals labeled O$p_x$ and O$p_y$ that form 
$\sigma$-bonds with the Cu$d$ orbital. The noninteracting tight-binding 
Hamiltonian in momentum space is given by
\begin{equation}
\hat H_0 = \sum_{\bk\sigma} 
\left(\begin{array}{ccc}\hat c^\dagger_{\bk d\sigma},&\hat c_{\bk x\sigma}^\dagger,
& \hat c_{\bk y\sigma}^\dagger,\end{array}\right)
{\bf H}_0(\bk) 
\left ( \begin{array}{c} 
\hat c_{\bk d\sigma} \\ \hat c_{\bk x\sigma} \\ \hat c_{\bk y\sigma}
\end{array}\right ),
\end{equation}
where $\hat c_{\bk\alpha\sigma}$ is the annihilation operator for an electron 
with wavevector ${\bk}$, spin $\sigma$, and orbital index $\alpha$; 
$\alpha= d,\,x,\,y$ indicates Cu$d$, O$p_x$, O$p_y$ orbitals, respectively.
In the matrix elements of 
\begin{eqnarray}
{\bf H}_0(\bk) &=&
\left (
\begin{array}{ccc}
	\epsilon_{d} & 2t_{pd} s_x & -2t_{pd} s_y \\
	2t_{pd} s_x & \epsilon_{x}		&   -4t_{pp} s_x s_y \\
	-2t_{pd} s_y & -4t_{pp} s_x s_y & \epsilon_{y}		
\end{array}
\right ),
\label{eq:H0}
\end{eqnarray}
with $s_x=\sin(k_x/2)$, $s_y=\sin(k_y/2)$, $t_{pd}$ and $t_{pp}$ are the nearest 
neighbor $p$-$d$ and $p$-$p$ hopping amplitudes (see Fig.~\ref{fig:unitcell}).
$\epsilon_d$ denotes the Cu$d$ orbital energy and $\epsilon_x$ and $\epsilon_y$ 
the corresponding energies of the O$p_x$ and O$p_y$ orbitals. 
 In calculating ${\bf H}_0$, the signs of the hopping matrix elements are determined by the phases  of the nearest atomic wavefunction lobes (indicated by ``$+$" and ``$-$" in Figure \ref{fig:unitcell}) for a given bond.
For the $H_{0,23}$ and $H_{0,32}$ matrix elements, we introduced an extra minus sign as that enables us to obtain a realistic Fermi surface for a non-zero $t_{pp}$ value.  Such a sign 
change can arise from indirect hopping through the Cu4s orbital.\cite{Andersen:1995}
Unless otherwise specified, we take $\epsilon_x=\epsilon_y\equiv\epsilon_p$, choose $t_{pd}=1$ as the unit of energy, and adopt the common estimate for the charge transfer energy
$\Delta_{CT}=\epsilon_d-\epsilon_p=2.5$\cite{Arrigoni:2009njp}. 
To understand the role of the Fermi-surface shape, we consider two cases: $t_{pp}=0$ and 
$t_{pp}=0.5$. The former gives a highly nested Fermi surface, while the latter 
is consistent with band-structure calculations for the 
cuprates.\cite{Hybertsen:1989prb}

The energy eigenvalues and eigenvectors follow from the diagonalization of 
${\bf H}_0(\bk)$,
\begin{equation}
{\bf S}^\dagger(\bk) {\bf H}_0(\bk) {\bf S}(\bk)={\bf \Lambda}(\bk),
\label{eq:Sk}
\end{equation}
where $\Lambda_{ij}(\bk)=\delta_{ij}E_\bk^i$ is the diagonal eigenvalue matrix 
containing the band energies  $E^i_\bk$, and ${\bf S}(\bk)$ is a $3\times 3$
matrix of eigenvectors.

The interaction term in the Hamiltonian includes intra- and inter-orbital 
Coulomb interactions. We consider intra-orbital interactions $U_d$ and $U_p$ at 
the Cu$d$ and O$p$ orbitals and inter-orbital interactions $V_{pd}$ and $V_{pp}$
between nearest neighbor $p$-$d$ and $p$-$p$ orbitals, respectively, as 
illustrated in Fig.~\ref{fig:unitcell}.

The interaction term is 
\begin{eqnarray}
\label{e:vaabb}
  \hat V
  &=& \frac{1}{2N}\sum_{\alpha\beta}\sum_{\bf q}V_{\alpha\beta}(\bq)
  \hat n_{\alpha}(\bq) \hat n_{\beta}(-\bq) 
\end{eqnarray}
where $\hat n_{\alpha}(\bq)=\sum_{\sigma \bk}
\hat c^\dagger_{\bk\alpha\sigma} \hat c^{}_{\bk+\bq\alpha\sigma}$ and
\begin{eqnarray}
V_{\alpha\beta}(\bq) 
&=& \sum_{\bf R}e^{-i\bq\cdot({\bf R} + {\bf r}_\beta-{\bf r}_\alpha)}
V_{\alpha\beta}({\bf R} ). 
\end{eqnarray}
${\bf R}$ denotes the lattice vectors and $\br_\alpha$,$\br_\beta$ the positions
of orbitals $\alpha$,$\beta$ within the unit cell. Explicitly,
\begin{eqnarray}
V_{\alpha\beta}(\bq) 
   &=& \delta_{\alpha, d}\delta_{\beta, d} U_d \\
   &+&\nonumber (\delta_{\alpha, x}\delta_{\beta, x}+\delta_{\alpha, y}\delta_{\beta, y}) U_p \\
   &+&\nonumber (\delta_{\alpha, x}\delta_{\beta, d}+\delta_{\alpha, d}\delta_{\beta, x}) 2V_{pd}\cos(q_x/2) \\
   &+&\nonumber (\delta_{\alpha, y}\delta_{\beta, d}+\delta_{\alpha, d}\delta_{\beta, y}) 2V_{pd}\cos(q_y/2)  \\
   &+& (\delta_{\alpha, y}\delta_{\beta, x}+ \delta_{\alpha, x}\delta_{\beta, y})
	4V_{pp}\cos(q_x/2)\cos(q_y/2). \nonumber 
\end{eqnarray}
Throughout this paper, we set $U_d=9$, $U_p=3$, and $V_{pd}=1$ (in units of 
$t_{pd}$).\cite{Hybertsen:1989prb}

\subsection{Charge Susceptibility}
\label{s:dpt}

In order to detect tendencies for nematic instabilities, we consider the order 
parameter
\begin{equation}
O_N(\bq)=n_x(\bq)-n_y(\bq),
\end{equation}
which measures the charge transfer between O$p_x$ and O$p_y$ orbitals. The 
corresponding  nematic susceptibility is given by
\begin{eqnarray}
	\chi_N(\bq) &=& \frac{\partial O_N(\bq)}{\partial \phi(\bq)} \\
	&=& \chi_{xx}(\bq)-\chi_{xy}(\bq)+\chi_{yy}(\bq)-\chi_{yx}(\bq).
	\label{eq:chiN}
\end{eqnarray}
$\phi(\bq)=\epsilon_x(\bq) -\epsilon_y(\bq)$ is a nematic perturbing potential,
and the charge susceptibility matrix is
\begin{equation}
\label{e:chi_intuit_def}
\chi_{\alpha\beta}(\bq) = \frac{\delta n_\alpha(\bq)}{\delta \epsilon_\beta(\bq)},
\end{equation}
with $\delta n_\alpha(\bq)$ the change in the charge density 
$n_\alpha(\bq)=\langle \hat n_{\alpha}(\bq) \rangle$ due to a weak perturbation
$\delta \epsilon_\beta(\bq)$ of the orbital energies. For dynamical 
perturbations with frequency $\omega$, Eq.~\reff{e:chi_intuit_def} generalizes 
to the dynamic susceptibility $\chi_{\alpha\beta}(\bq,\omega)$. For the 
remainder of this section, we adopt the shorthand notation 
$q\equiv (\bq,\omega)$.

\begin{figure}[tb]
 \includegraphics[width=\columnwidth]{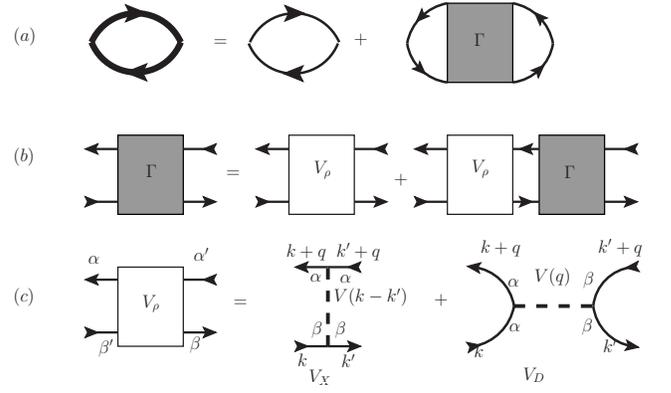}
 \caption{ (a) Diagrammatic representation of the dynamic charge susceptibility
    Eq.~\reff{e:chiladder} in terms of the bare susceptibility and the 
    interaction vertex $\Gamma(\bk,\bk^\prime,q)$. (b) Ladder diagrams 
    corresponding to Eq. \reff{e:gamma} for the interaction vertex. (c) The 
    effective interaction in the charge channel, as in Eq.~\reff{eq:vrho}.  }
  \label{f:bethe-salp}
\end{figure}

The dynamic charge susceptibility is evaluated using the Kubo formula and an 
infinite summation of ladder and bubble diagrams, as shown in 
Fig.~\ref{f:bethe-salp}. For this purpose it is useful to define an effective 
interaction for the charge channel,\cite{Littlewood:1990} as represented 
diagrammatically in Fig. \ref{f:bethe-salp}(c),
\begin{eqnarray}
	V_{\rho,\alpha\alpha^\prime\beta\beta^\prime}(\bk,\bk^\prime,\bq)
        &=& -\delta_{\alpha^\prime, \alpha}\delta_{\beta^\prime,
          \beta} V_{\alpha\beta}(\bk^\prime -\bk) \nonumber
        \\ &&+\delta_{\alpha^\prime, \beta}\delta_{\beta^\prime,
          \alpha}2V_{\alpha\beta}(\bq) .
	\label{eq:vrho}
\end{eqnarray}
The first and the second term on the right describe the exchange and the
direct interaction, respectively. Similarly, for the spin susceptibility the 
same set of diagrams is evaluated using the effective interaction in the spin 
channel,\cite{Littlewood:1990}
\begin{eqnarray}
	V_{\sigma,\alpha\alpha^\prime\beta\beta^\prime}(\bk,\bk^\prime,\bq)
        &=& -\delta_{\alpha^\prime, \alpha}\delta_{\beta^\prime,
          \beta} V_{\alpha\beta}(\bk^\prime -\bk).
		\label{eq:vsigma}
\end{eqnarray}
With Eq.~\reff{eq:vrho}, the interaction vertex 
$\Gamma_{\alpha\alpha^\prime\beta\beta^\prime}(\bk,\bk^\prime,q)$ in the charge 
channel is conveniently obtained in the compact form shown in 
Fig.~\ref{f:bethe-salp}(b).  Once this equation is solved for
$\Gamma_{\alpha\alpha^\prime\beta\beta^\prime}(\bk,\bk^\prime,q)$, the
susceptibility is calculated from the diagrams in Fig.~\ref{f:bethe-salp}(a).

In order to solve the integral equation for the interaction vertex we project 
onto a set of basis functions in orbital and momentum 
space,\cite{Littlewood:1989} which transforms the integral equation in
Fig.~\ref{f:bethe-salp}(b) into a matrix equation:
\begin{eqnarray}
	V_{\alpha\beta}(\bk^\prime-\bk) &=& \sum_{i,j}
        g^i_{\alpha\beta}(\bk) \tilde V^{ij}_{X}
        g^j_{\alpha\beta}(\bk^\prime), \label{e:vsepareted}
        \\ V_{\alpha\beta}(\bq) &=& \sum_{i,j} g^i_{\alpha\alpha}
        \tilde V^{ij}_{D}(\bq) g^j_{\beta\beta},
	\label{e:vprod} \\
\Gamma_{\alpha\alpha'\beta\beta'}(\bk,\bk^\prime,q) &=& \sum_{i,j} 
g^i_{\alpha\beta'}(\bk) \tilde \Gamma^{ij}(q) g^j_{\alpha'\beta}(\bk^\prime). 
\label{e:gamproj}
\end{eqnarray}
The functions $g^i_{\alpha\beta}(\bk)$ form a 19-dimensional basis 
($i\in[1,19]$), and they are explicitly defined in Appendix \ref{s:fbasis}. 
${\bf \tilde X}$ denotes the matrix representation, with matrix elements 
$\tilde X^{ij}$, of a quantity $X$ with respect to the basis functions. In this 
notation ${\bf \tilde V}_D(\bq)$, ${\bf \tilde V}_X$, and 
${\bf \tilde \Gamma}(q)$ are defined by 
Eqs.~\reff{e:vsepareted}--\reff{e:gamproj}. ${\bf \tilde V}_D(\bq)$ and 
${\bf \tilde V}_X$ are also explicitly given in Appendix \ref{s:fbasis}.
We note that $\tilde V_D^{ij}(\bq)$ is nonzero only for $i,j = 9,10,11$, for 
which $g^i_{\alpha\beta}(\bk)$ does not explicitly depend on $\bk$.

Using Eqs.~\reff{e:vsepareted} and \reff{e:vprod}, we obtain
\begin{equation}
	\tilde {\bf V}_\rho(\bq) = 2\tilde {\bf V}_D(\bq) - \tilde {\bf V}_X.
\end{equation}
This equation, combined with Eq. \reff{e:gamproj}, enables one to express the integral
equation in Fig.~\ref{f:bethe-salp}(b) as a matrix equation, the inversion of 
which leads to
\begin{equation}
{\tilde {\bf \Gamma}}(q) = [{\bf 1}+{\tilde {\bf V}_\rho(\bq)} 
\tilde \chibf_0(q)]^{-1}{ \tilde {\bf V_\rho}(\bq)}.
\label{e:gamma}
\end{equation} 
$\tilde\chibf_0(q)$ is the projected bare susceptibility,
\begin{eqnarray}
  \tilde \chi_0^{ij}(q) \nonumber 
 &=&\frac{-1}{N} \sum_{\bk,\mu\nu}\sum_{\theta\theta^\prime\gamma \gamma^\prime} 
 g^i_{\gamma^\prime\gamma}(\bk) M^{\gamma^\prime\theta\theta^\prime\gamma}_{\nu\mu \bk \bq} \\
 & &	\hspace{0.08\columnwidth}\times  
	\frac{[f(E^{\nu}_{\bk})-f(E^{\mu}_{\bk+\bq})]}{\omega+E^{\nu}_{\bk} - E^{\mu}_{\bk+\bq}+{\rm i}\delta}
		g^j_{\theta\theta^\prime}(\bk),
\label{e:P0}
\end{eqnarray}
with $f(E)$ the Fermi function and 
\begin{eqnarray}
	M^{\gamma^\prime\theta\theta^\prime\gamma}_{\nu\mu \bk \bq} 
	= S_{\gamma^\prime\nu}(\bk) S_{\theta\nu}^*(\bk)
	S_{\theta^\prime\mu}(\bk+\bq)S^*_{\gamma\mu}(\bk+\bq).
\end{eqnarray}
The charge susceptibility then follows as
\begin{eqnarray}
 \chi_{\alpha\beta}(q) 
	&=& \chi^0_{\alpha\beta}(q)  - \sum_{ij} A_{\alpha\alpha}^i(q)\tilde{\bf \Gamma}^{ij}(q)A_{\beta\beta}^j(q),
   \label{e:chiladder}
\end{eqnarray}
where 
\begin{eqnarray}
	\chi_{\alpha\beta}^0(q)
	&=&	\frac{-1}{N} \sum_{\bk,\mu\nu} 
		M^{\beta\alpha\alpha\beta}_{\nu\mu \bk \bq}   
		\frac{[f(E^{\nu}_{\bk})-f(E^{\mu}_{\bk+\bq})]}{\omega+E^{\nu}_{\bk} - E^{\mu}_{\bk+\bq}+{\rm i}\delta}
\label{e:C0}
\end{eqnarray}
is the bare charge susceptibility, and 
\begin{eqnarray}
	A^{j}_{\gamma\gamma^\prime}(q) 
	&=&	\frac{1}{N} \sum_{\bk,\mu\nu}\sum_{\theta\theta^\prime}  
		M^{\gamma^\prime\theta\theta^\prime\gamma}_{\nu\mu \bk \bq}   
		\frac{f(E^{\nu}_{\bk})-f(E^{\mu}_{\bk+\bq})}{\omega+E^{\nu}_{\bk} - E^{\mu}_{\bk+\bq}+{\rm i}\delta}
		g^j_{\theta\theta^\prime}(\bk). \nonumber \\
\label{e:ajgg}
\end{eqnarray}
Equation~\reff{e:chiladder}, which corresponds to Fig.~\ref{f:bethe-salp}(a), 
is the final result for the charge susceptibility.

\section{Results}
\label{sec:results}
We identify charge instabilities from divergences of the charge susceptibility 
matrix. The focus is on nematic order that is marked by a divergence of 
$\chi_N(\bq)$ and driven by the repulsive interaction $V_{pp}$. We start with a 
general discussion of MNO and the structure of $\chi_N(\bq)$ and then present 
the results which show how the instability depends on $V_{pp}$, the hole density
$p=5-n$, and the temperature $T$. $n$ is the number of electrons per unit cell, and
$p$ is thus measured relative to half-filling of the Cu$d$ orbital. We consider 
the two cases $t_{pp}=0$ and $t_{pp}=0.5$, which correspond to a small and a 
large Fermi surface curvature, respectively. The latter value is motivated by 
band-structure calculations and matches reasonably well the Fermi surface of 
cuprate materials. Subsequently we discuss the spectrum of the dynamical 
susceptibility $\chi_N(\bq,\omega)$ in the isotropic phase near the nematic 
transition and the consequences of MNO for the reconstruction of the Fermi 
surface. 

\subsection{Nematic Susceptibility}

\begin{figure}[tb]
\includegraphics[width=\columnwidth]{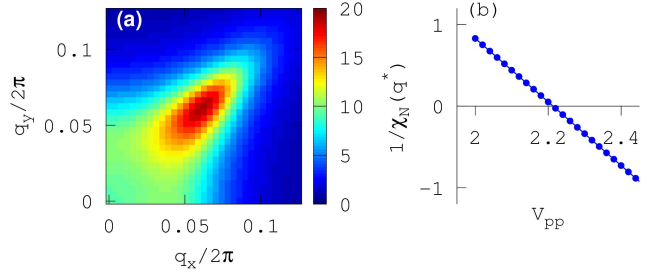}
  \caption{(a) Nematic susceptibility $\chi_N(\bq)$ for $p=0.15$, $T=0.016$, 
    $V_{pp}=2.2$, and $t_{pp}=0.5$. (b) $V_{pp}$ dependence of 
    $1/\chi_N(\bq^\ast)$ showing the divergence of the susceptibility close to
    $V_{pp} = 2.22$.  }
\label{f:chiq}
\end{figure}

As an example, we show in Fig.~\ref{f:chiq}(a) the static nematic 
susceptibility $\chi_N(\bq)$ for a set of parameters near the MNO instability. 
For the selected parameters $\chi_N(\bq)$ is sharply peaked at the diagonal 
wavevector ${\bq^\ast}/2\pi=0.06(1,1)$.  The inverse susceptibility at 
${\bq^\ast}$ is plotted in Fig.~\ref{f:chiq}(b) as a function of $V_{pp}$; this 
figure shows that $\chi_N(\bq^\ast)$ diverges at $V_{pp} = 2.22$, which 
marks the boundary between an isotropic phase for $V_{pp}<2.22$ and  MNO 
with a modulation wavevector $\bq^\ast$.   In fact, because
 $\chi_N(\bq)$ has the full point group symmetry of the lattice, it actually
 diverges simultaneously at four distinct $\bq$-values related to $\bq^\ast$ by $\pi/2$ rotations.
Our calculations therefore admit both unidirectional order (involving only $\bq^\ast$ and $-\bq^\ast$) and bidirectional  checkerboard order (involving all four $\bq$ vectors).   Further extensions of the calculations are required to determine which phase is dominant, and in single-band models it has been shown that this depends on details of the interaction and Fermi surface.\cite{Melikyan:2011}

\begin{figure}[tb]
\includegraphics[width=\columnwidth]{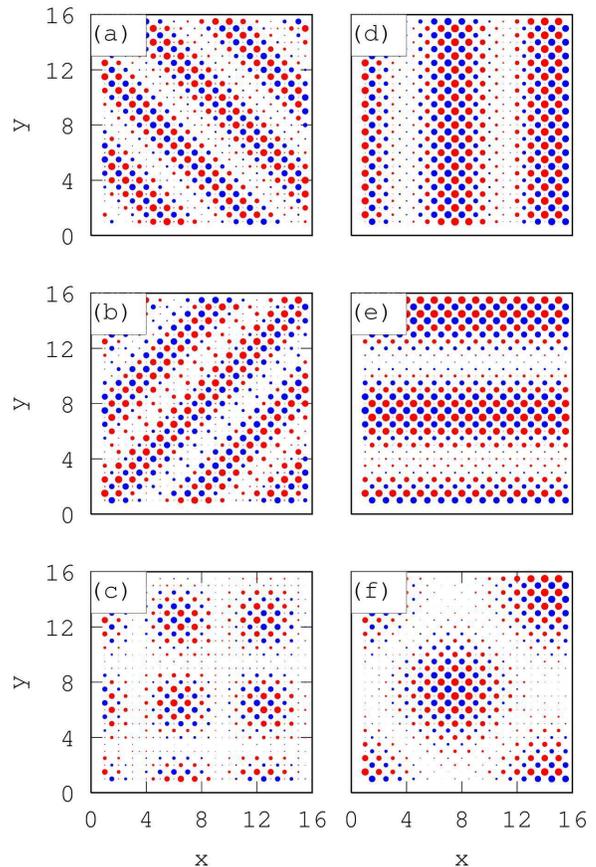}
\caption{ Charge density modulations for (a)-(c) diagonal and (d)-(f)
 axial modulated nematic order, at hole densities $p=0.14$ and $p=0.20$,
 respectively, for $t_{pp}=0.5$.  The corresponding $q^\ast$ values are given 
 in Fig.~\ref{fig:phaseboundary2} (b). Circle diameter indicates the magnitude 
 of the charge modulation, with red (blue) corresponding to a negative 
 (positive) modulation. While all orbitals are shown, modulations are only 
 large enough to be seen on the oxygen $p_x$ and $p_y$ orbitals. The 
 bidirectional patterns in (c) are obtained by adding the patterns in (a) and 
 (b); those in (f) are obtained by adding (d) and (e).}
\label{fig:CDWpattern}
\end{figure}

Depending on model parameters, three kinds of nematic phases emerge from the 
calculations: a commensurate phase ($\bq^\ast={\bf 0}$), the diagonal phase 
with $\bq^\ast = (q_0,q_0)$ as described above, and an axial phase for which 
the modulation wavevector is aligned with the crystalline axes, i.e.  
$\bq^\ast=(q_0,0)$ or $\bq^\ast=(0,q_0)$. The charge modulations associated with
both the diagonal and the axial phases are illustrated in 
Fig.~\ref{fig:CDWpattern}. These figures are for model parameters near the MNO 
phase boundary, where the response to a weak perturbing potential is large. We 
show results for unidirectional and bidirectional nematic modulations: for the 
unidirectional case, the charge modulations are generated assuming a nematic 
perturbing potential of the form $\delta\epsilon_d=0$, $\delta\epsilon_x(\br)= 
-\delta\epsilon_y(\br)=\delta\epsilon\cos(\bq^\ast \cdot {\bf r})$, for which
\begin{equation}
\delta n_\alpha(\br) = \sum_\beta \chi_{\alpha\beta} (\bq^\ast) \delta\epsilon_\beta(\br).
\label{eq:delta_n}
\end{equation}
For the bidirectional case, a second perturbation, with $\bq^\ast$ rotated by 
$\pi/2$, is added to the right hand side of Eq.~\reff{eq:delta_n}.

Figure~\ref{fig:CDWpattern} shows that in all cases, the charge transfer occurs
almost entirely between the oxygen atoms; the Cu$d$ charge modulations are at 
least an order of magnitude smaller than the oxygen modulations 
and are too small to see in the figure. Furthermore, in each of the MNO patterns
shown in Fig.~\ref{fig:CDWpattern} the charge transfer is predominantly within
the unit cell, with the inter-unit cell charge transfers being orders of 
magnitude smaller.  For this reason, it is  natural to think of the  broken 
symmetry phase as a modulated nematic rather than a CDW.

\subsection{Phase Boundaries}
\begin{figure}[tb]
\includegraphics[width=\columnwidth]{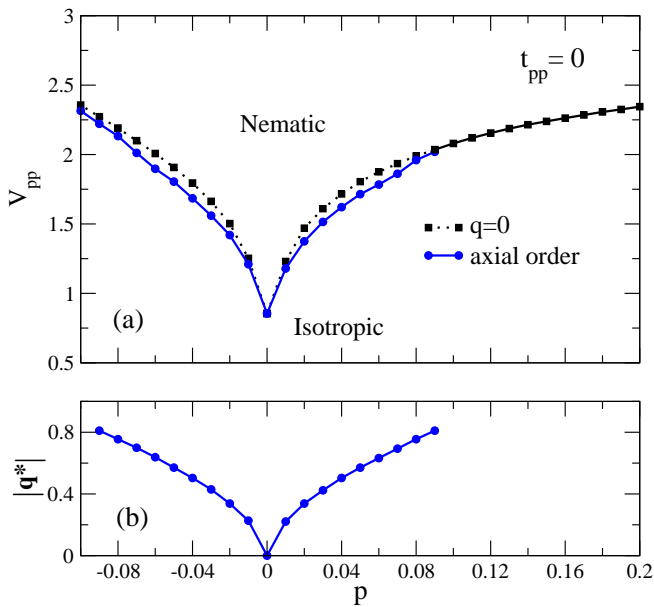}
\caption{(a) Charge instabilities for the commensurate ($\bq={\bf 0}$) and
  axial modulated nematic phases for $t_{pp}=0$ at $T=0.0005$. The system is
  isotropic for small $V_{pp}$. Solid curves indicate the leading divergence of 
  $\chi_N(\bq)$ as $V_{pp}$ is increased; dashed curves indicate subleading 
  instabilities. The phase transition is thus to an axial nematic phase for 
  $p\leq 0.09$, and to a commensurate nematic phase for $p>0.09$. The 
  modulation wavevector for the axial nematic phase is shown in (b), and it
  vanishes at the van Hove filling $p_\mathrm{vH}=0.0$. The model parameters are
  $t_{pd}=1$, $\epsilon_d=0$, $\Delta_{CT}=\epsilon_d-\epsilon_p=2.5$, 
  $V_{pd}=1$, $U_d=9$, and $U_p=3$.  }
\label{fig:phaseboundary1}
\end{figure}

\begin{figure}[tb]
\includegraphics[width=\columnwidth]{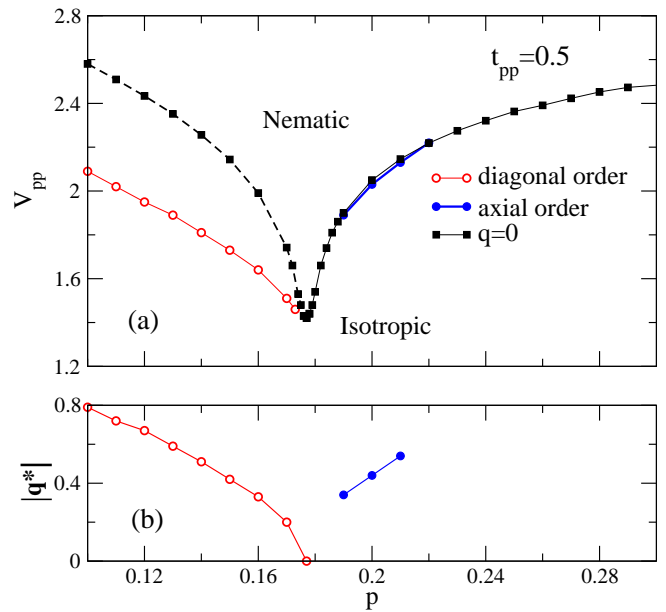}
\caption{As in Fig.~\ref{fig:phaseboundary1}, but for $t_{pp} = 0.5$. Here, the 
leading instability is to a diagonal nematic phase for $p<p_\mathrm{vH}$, where 
$p_\mathrm{vH}=0.177$ for this value of $t_{pp}$.}
  
\label{fig:phaseboundary2}
\end{figure}
We now examine how the phase boundary between isotropic and nematic phases 
depends on various model parameters. To begin with, we show in 
Figs.~\ref{fig:phaseboundary1}(a) and \ref{fig:phaseboundary2}(a) the phase 
boundaries at low $T$ in the $p$-$V_{pp}$ plane for $t_{pp}=0$ and $t_{pp}=0.5$, 
respectively.  In both figures, the system is isotropic for small $V_{pp}$, and 
the phase boundary marks where $\chi_N(\bq^\ast)$ first diverges as $V_{pp}$ is 
increased. The value of $\bq^\ast$ at which this happens depends on the hole 
density $p$, and is shown in Figs.~\ref{fig:phaseboundary1}(b) and
\ref{fig:phaseboundary2}(b).

Figures \ref{fig:phaseboundary1}(a) and \ref{fig:phaseboundary2}(a) also include
the curves along which $\chi_N(\bq={\bf 0})$ diverges. We see that there is a 
wide range of $p$ over which $\chi_N(\bq^\ast)$ diverges at a lower $V_{pp}$ 
than $\chi_N({\bf 0})$, signalling that MNO dominates over commensurate nematic 
order. There are also regions where only the $\bq={\bf 0}$ instability is shown;
in these regions this is the first instability which appears as $V_{pp}$ is 
increased.

The magnitude and the orientation of $\bq^\ast$ depend on the filling relative 
to the ``van Hove filling'' $p_\mathrm{vH}$, which is defined as the hole 
density $p$ at which the van Hove singularities at $(\pi,0)$ and $(0,\pi)$ cross
the Fermi energy. $p_\mathrm{vH}$ marks the boundary between hole-like 
($p<p_\mathrm{vH}$) and electron-like ($p>p_\mathrm{vH}$) Fermi surfaces. The 
value of $p_\mathrm{vH}$ depends on the Fermi surface curvature: 
$p_\mathrm{vH}=0$ for $t_{pp}=0$ and $p_\mathrm{vH}=0.177$ for $t_{pp}=0.5$. For 
reference, cuprate superconductors have hole-like Fermi surfaces in the doping 
range where charge ordered phases are observed.

In both Figs.~\ref{fig:phaseboundary1} and \ref{fig:phaseboundary2}, the 
nematic instability is to a commensurate phase at large $p$. At lower fillings,
but still above $p_\mathrm{vH}$, the leading instability is incommensurate, 
with an axial modulation wavevector that decreases as $p$ is reduced: when
$t_{pp}=0$, $\bq^\ast$ vanishes continuously as $p\rightarrow p_{vH}$, at which 
point the charge order is commensurate. For $t_{pp}=0.5$, there is a 
discontinuous transition to the commensurate phase at $p=0.19>p_\mathrm{vH}$.  
Below the van Hove filling, $|\bq^\ast|$ grows as $p$ is reduced. In this range
of doping the modulation wavevector is axial for $t_{pp}=0$
(Fig.~\ref{fig:phaseboundary1}) and diagonal for $t_{pp}=0.5$ 
(Fig.~\ref{fig:phaseboundary2}). For the relative hole filling $p=p_{vH}-0.05$, the crossover between diagonal 
and axial order occurs for $t_{pp}\approx0.12$. As we discuss further in Sec.~\ref{sec:discussionA}, the overall 
doping dependence is similar to that found by Holder and 
Metzner.\cite{Holder:2012ks}
 
 \begin{figure}[tb]
\includegraphics[width=\columnwidth]{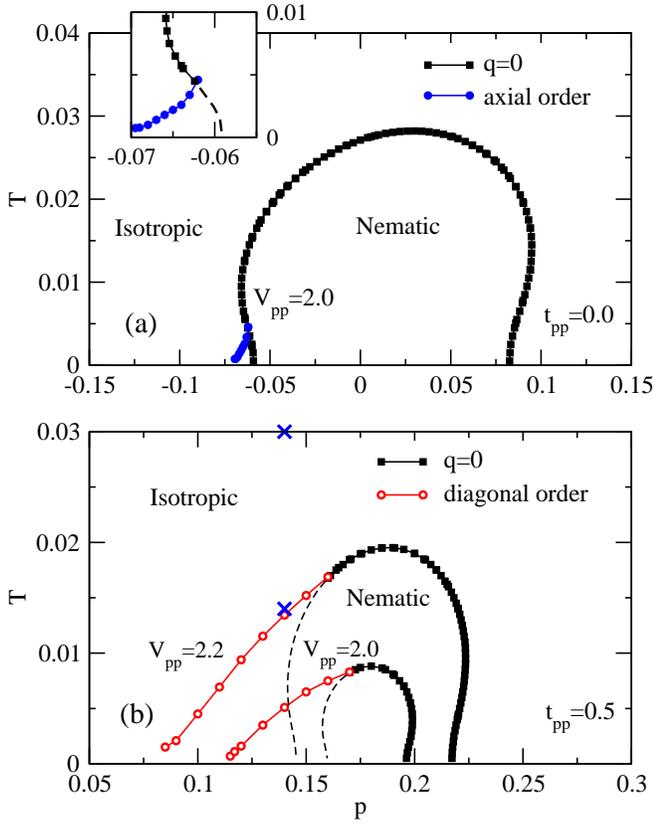}
\caption{Phase diagram in the $p$-$T$ plane for (a) $t_{pp}=0$ and (b) 
  $t_{pp}=0.5$. Solid curves indicate the leading divergence of $\chi_N(\bq)$ 
  upon cooling; dashed lines indicate subleading instabilities. The inset in (a)
  shows a zoom into the left corner of the nematic region where axial order is 
  dominant. The crosses in (b) indicate the points in the phase diagram at 
  which the dynamic susceptibility is shown in Fig.~\ref{f:chiw}. Model
  parameters are as in Fig.~\ref{fig:phaseboundary1}.  }
\label{fig:braintpp0.5}
\end{figure}
 
In Fig.~\ref{fig:braintpp0.5} we map the nematic instability in the $p$-$T$ 
plane for fixed $V_{pp}$. Results for $t_{pp}=0$ and $t_{pp}=0.5$ are shown in 
Figs.~\ref{fig:braintpp0.5}(a) and (b), respectively. In both cases, the system
is isotropic at high $T$, and the leading instability upon cooling is indicated
by a solid line; dashed lines indicate subleading instabilities.  

Figure \ref{fig:braintpp0.5}(a) is dominated by a transition to commensurate 
nematic order extending over a broad doping range that includes both hole-like 
($p<0$) and electron-like ($p>0$) Fermi surfaces. The system exhibits 
re-entrance at both the lower and upper ends of the doping range where nematic 
order is encountered: as $T$ is lowered, the system passes through an 
isotropic-to-nematic transition followed by a nematic-to-isotropic transition. 
The shape of the commensurate phase boundary is essentially the same as that 
found in a previous mean-field study.\cite{Fischer:2011} The new result in
Fig.~\ref{fig:braintpp0.5}(a) is the existence of a phase boundary spanning 
$-0.07 < p < -0.062$ between isotropic and axial MNO phases at low $T$. This 
figure shows that the axial MNO phase found at low $T$ 
(Fig.~\ref{fig:phaseboundary1}) is fragile for $t_{pp}=0$ and subdominant to
the commensurate phase at higher $T$.

While the MNO phase boundary represents only a small fraction of
Fig.~\ref{fig:braintpp0.5}(a), it is much more important when we adopt
the more realistic value of $t_{pp}=0.5$ in
Fig.~\ref{fig:braintpp0.5}(b). In particular, the leading instability
is to a diagonal MNO phase for a wide doping range of hole-like Fermi
surfaces ($p<p_\mathrm{vH}$). This doping range is similar to the
range over which charge modulations are observed
experimentally. However, there are two significant discrepancies with
the experiments: the values of $\bq^\ast$ obtained here are a factor
of 2 to 3 smaller than those observed experimentally, and the
orientation of $\bq^\ast$ is diagonal, while experiments observe axial
order. The first discrepancy might be resolved by using a band
structure tailored specifically to ortho-II YBCO;\cite{Carrington:2007}
this likely requires going beyond the three-band model.
These issues will be discussed in more detail in Sec.~\ref{sec:discussionB}. 

Finally, we note that the nematic instability is actually subleading
for our model parameters; the leading instability is to a spin-density
wave (SDW) state.  A calculation of the spin susceptibility, performed
in the same manner as for the charge susceptibility, but using the
effective interaction in the spin channel Eq.~\reff{eq:vsigma} in
place of Eq.~\reff{eq:vrho}, shows that the SDW involves primarily the
Cu$d$ orbitals and is driven by the large on-site repulsion $U_d$ on
the copper atoms. For $p=0.14$, $t_{pp}=0.5$, and $V_{pp}=2.2$, the
onset temperature for the SDW is higher than the nematic transition
temperature provided $U_d > 1.08$ while for smaller $U_d$, the nematic
instability occurs at higher $T$. There are two reasons why this SDW
phase may not matter for nematic order.  First, strong correlation
effects, neglected here, will renormalize the interaction vertex, and
thereby lead to a lower magnetic transition temperature than predicted
by weak-coupling theory.  Second, the onset of SDW order does not
preclude nematic order becase the SDW and MNO involve different
oribitals.  This raises the intriguing possibility that these phases
might coexist, with little or no competition, in some regions of the
phase diagram.
 

\subsection{Dynamical Susceptibility}
\begin{figure}[tb]
\includegraphics[width=\columnwidth]{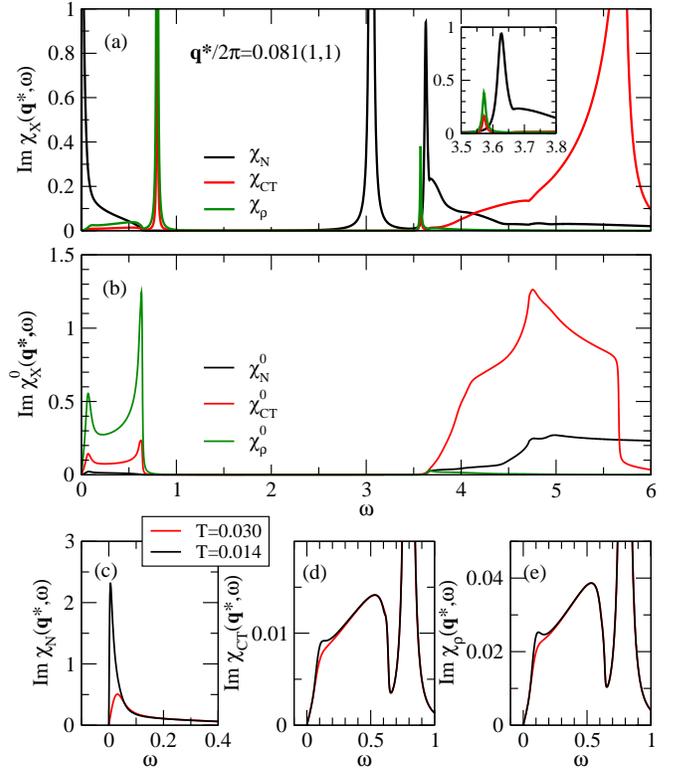}
  \caption{Imaginary part of the frequency dependent susceptibility at 
$\bq^\ast$. Results are for $t_{pp}=0.5$, hole density $p=0.14$,
and $V_{pp}=2.2$. (a) Plots of the nematic, the charge-transfer, and
the charge susceptibility slightly above the nematic transition
($T=0.014$). The inset shows an expanded view of the frequency range
$3.5 <\omega<3.8$. For comparison, the bare susceptibilities are shown
in (b).  The temperature dependence of the low-frequency spectrum is
separately displayed for the nematic (c), the charge-transfer (d) and
the charge susceptibility (e). Panel (c) shows the emergence of a
sharp low-frequency peak in the nematic susceptibility near the
nematic transition.  For reference, the points on the phase diagram at
which the susceptibilities are evaluated are marked with crosses in
Fig.~\ref{fig:braintpp0.5}(b). }
\label{f:chiw}
\end{figure}

Figure~\ref{f:chiw} shows the dynamical susceptibility at the modulation 
wavevector $\bq^\ast$ in the isotropic phase. Results are shown for the same 
parameters as in Fig.~\ref{fig:braintpp0.5}(b), with $p=0.14$, $V_{pp}=2.2$, and
for $T=0.014$ and $T=0.030$. These filling and temperature values are indicated 
by crosses in Fig.~\ref{fig:braintpp0.5}(b).
The lower temperature, $T=0.014$, is close to 
the nematic transition while the higher temperature, $T=0.030$, is 
approximately twice the transition temperature.  

Besides the nematic susceptibility $\chi_N(\bq,\omega)$, the total charge 
susceptibility
\begin{eqnarray}
\chi_{\rho}(\bq,\omega)=
\frac{\partial(n_d+n_x+n_y)}{\partial(\epsilon_d+\epsilon_x+\epsilon_y)}= 
\sum_{\alpha,\beta} \chi_{\alpha\beta}(\bq,\omega) 
\end{eqnarray}
and the charge-transfer susceptibility\cite{Littlewood:1990}
\begin{eqnarray}
\chi_\mathrm{CT}(\bq,\omega) &=& 
\frac{\partial(n_d-n_x-n_y)}{\partial(\epsilon_d-\epsilon_x-\epsilon_y)} \\
&=&\chi_\rho(\bq,\omega) - 
2\sum_{\alpha=x,y} \left[ \chi_{d\alpha}(\bq,\omega) + \chi_{\alpha d}(\bq,\omega)
\right ]. \nonumber 
\end{eqnarray}
are included. The frequency dependence of these three susceptibilities are 
shown in Fig.~\ref{f:chiw}(a) in comparison to the bare versions of these 
susceptibilities in Fig.~\ref{f:chiw}(b). The bare spectra consist of a 
low-energy intraband continuum due to particle-hole excitations with momentum 
$\bq^\ast$, and a high-energy interband continuum. The width of the low-energy 
continuum is ${\bf q}$-dependent, and vanishes as $q\rightarrow 0$; the 
interband particle-hole continuum is instead only weakly ${\bf q}$-dependent.  

The particle-hole continuum is renormalized by the interactions. In particular, 
the low-frequency charge susceptibility $\chi_\rho(\bq^\ast,\omega)$ is reduced 
by an order of magnitude relative to $\chi_\rho^0(\bq^\ast,\omega)$. This 
originates from the large value of the on-site Coulomb interaction, $U_d=9$, 
which suppresses charge fluctuations on the Cu$d$ orbitals. The
susceptibilities in Fig.~\ref{f:chiw}(a) have a number of additional 
resonances. Just above the low-frequency continuum, at $\omega=0.8$, both 
$\chi_\mathrm{CT}(\bq^\ast,\omega)$ and $\chi_\rho(\bq^\ast,\omega)$ exhibit a 
sharp resonance, which was identified before as a zero-sound 
mode.\cite{Bang:1993} It is this mode, rather than the charge-transfer 
excitation, which becomes soft at the charge-transfer instability. However, 
this mode is not relevant for the nematic transition as it remains at finite 
frequency across the transition.

The second pronounced resonance, at $\omega=5.58$, is the charge-transfer 
excitation. This mode corresponds to a transfer of charge between Cu$d$ and 
O$p$ orbitals without any change in the total intra-unit cell charge density;
consequently, the peak appears in $\chi_\mathrm{CT}(\bq^\ast,\omega)$ but not in 
$\chi_\rho(\bq^\ast,\omega)$. A third excitonic resonance appears in both
$\chi_\mathrm{CT}(\bq^\ast,\omega)$ and $\chi_\rho(\bq^\ast,\omega)$ at 
$\omega=3.57$, just below the high-frequency interband continuum. As shown in 
the inset, this excitonic peak is distinct from a nearby nematic resonance at 
$\omega=3.63$. All three modes (zero sound, excitonic, and charge transfer) have
$A_{1g}$ symmetry as $q\rightarrow 0$. In contrast, resonances in the nematic
susceptibility have $B_{1g}$ symmetry. This is the same symmetry as for the 
$d$-wave Pomeranchuk instability.\cite{Yamase:2004,Yamase:2005}

Three resonances are visible in the nematic susceptibility. There are two 
excitonic resonances at $\omega=3.63$ and $\omega = 3.05$ and a low-frequency 
peak that becomes soft at the nematic transition. All three excitations involve 
only O$p_x$ and O$p_y$ orbitals, such that there is no peak in
$\chi_{\alpha\beta}(\bq^\ast,\omega)$, if either $\alpha=d$ or $\beta=d$.  

As shown in Fig.~\ref{f:chiw}(c)-(e) the low-frequency spectra of both 
$\chi_\rho(\bq^\ast,\omega)$ and $\chi_\mathrm{CT}(\bq^\ast,\omega)$ are only 
weakly $T$-dependent. The low-frequency nematic mode, on the other hand, depends
strongly on $T$. In particular, it is damped by the particle-hole continuum away
from the nematic transition, but sharpens significantly as $T$ is lowered, and 
the excitation frequency shifts towards zero. The approach to the nematic 
transition is therefore accompanied by the softening of a mode that emerges 
from the particle-hole continuum.

\subsection{Fermi Surface in the MNO Phase}
\label{sec:FermiSurface}
\begin{figure}[tb]
\includegraphics[width=\columnwidth]{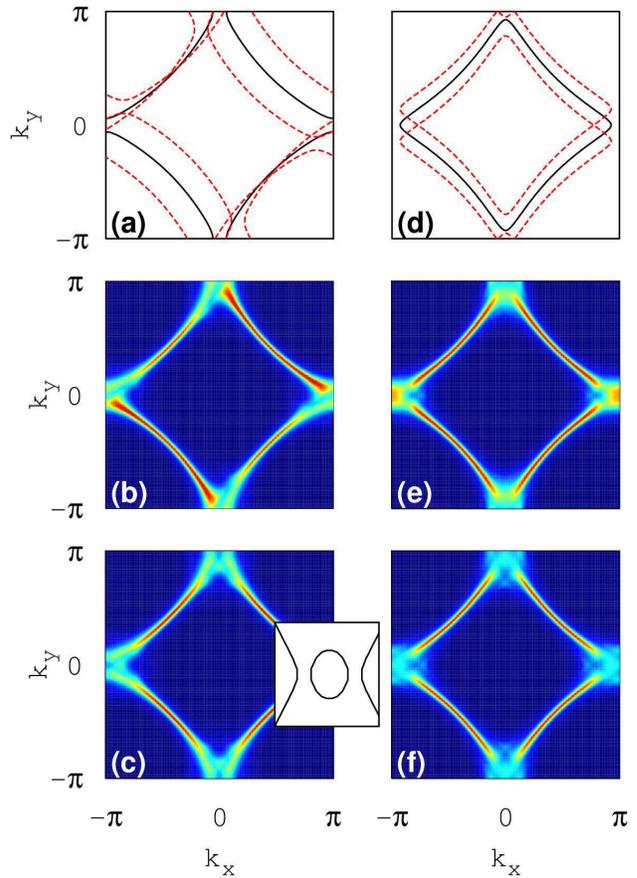} 
\caption {Fermi surface reconstruction for (a)-(c) diagonal ($p=0.14$) and 
  (d)-(f) axial ($p=0.20$) modulation vectors $\bq^\ast$ for $t_{pp}=0.5$. 
  Solid black lines in (a) and (d) show Fermi surface contours, and dotted red 
  lines show the same contours shifted by $\pm \bq^\ast$. Intensity plots in 
  (b),(c),(e),(f) show the spectral function, $A(\bk,\varepsilon_F)$. Results 
  are for (b),(e) unidirectional and (c),(f) bidirectional modulations. 
  In (c), the inset shows a zoom of the reconstructed Fermi surface, centered 
  at $(\pi,0)$. In all panels, $\bq^\ast$ is taken from 
  Fig.~\ref{fig:phaseboundary2} for the corresponding filling. A Lorentzian 
  broadening of $\delta = 0.05$ was used to calculate the spectral functions.}
\label{fig:FS_reconstruction}
\end{figure}
To understand the effects of modulated nematic order on the single-particle 
spectrum, we recalculate the band structure of the three-band model in the
presence of nematic order $\delta\epsilon_x(\br)=-\delta\epsilon_y(\br)=
\delta\epsilon\cos(\bq^\ast\cdot\br)$, in the same fashion used to generate 
Fig.~\ref{fig:CDWpattern}. For the unidirectional modulations, we start from 
an approximate perturbed Hamiltonian of the form
\begin{equation}
{\bf H}(\bk) = \left [ \begin{array}{ccc}
{\bf H}_0(\bk) & {\bf H}_{\bq^\ast}&  {\bf H}_{-\bq^\ast} \\
 {\bf H}_{\bq^\ast} & {\bf H}_0(\bk+\bq^\ast) & 0 \\
 {\bf H}_{-\bq^\ast} & 0 & {\bf H}_0(\bk-\bq^\ast) 
\end{array} \right ],
\label{eq:H9}
\end{equation}
where ${\bf H}_0(\bk)$ is the same $3\times 3$ matrix as in
Eq.~\reff{eq:H0}.  ${\bf H}_{\pm
  \bq^\ast}=(\delta\epsilon/2)\mbox{diag}(0,1,-1)$ is the matrix
representation of the perturbing potential, which scatters
quasiparticles by $\pm \bq^\ast$. ${\bf H}(\bk)$ is a $9\times 9$
matrix that is diagonalized numerically for each $\bk$.  In principle,
incommensurate nematic order also mixes in states with momenta $\bk\pm
2\bq^\ast$, $\bk \pm 3\bq^\ast$, etc.. However, we find that these
states rapidly diminish in importance and that the spectral function
can be understood in terms of the leading order terms alone.
 
The Fermi surfaces are obtained from the eigenvalues of ${\bf H}(\bk)$, while 
the spectral function is taken from the trace over Cu$d$, O$p_x$, and O$p_y$ 
orbitals, namely
\begin{equation}
A(\bk,\omega)=-\frac{\mbox{Im }}{\pi}\sum_{i=1}^3\left[(\omega+{\rm i}\delta) 
{\bf 1} - {\bf H}(\bk) \right ]^{-1}_{ii}
\end{equation}
where $[\ldots]^{-1}_{ii}$ is the $i$th diagonal element of the matrix inverse. 
The Fermi surfaces and the spectral functions are plotted in 
Fig.~\ref{fig:FS_reconstruction} in the first Brillouin zone. The same two 
cases are considered here as in Fig.~\ref{fig:CDWpattern}: 
$p=0.14<p_\mathrm{vH}$, for which $\bq^\ast$ is diagonal, and 
$p=0.20>p_\mathrm{vH}$, for which $\bq^\ast$ is axial; the values for $\bq^\ast$ 
are taken from Fig.~\ref{fig:phaseboundary2}.

In Figs.~\ref{fig:FS_reconstruction}(a) and (d), we show contours of the bare 
Fermi surface, along with Fermi-surface replicas shifted by $\pm \bq^\ast$. In 
Fig.~\ref{fig:FS_reconstruction}(a), the bare and shifted Fermi surface segments
coincide near $(\pi,0)$ and $(0,\pi)$ (the antinodes in the language of 
$d$-wave superconductivity), which
suggests that the diagonal nematic modulation is driven by nesting of short 
Fermi surface segments in the antinodal regions. Nesting features are not 
obvious in Fig.~\ref{fig:FS_reconstruction}(d), but there is a short segment of
the Fermi surface near $(0,\pi)$ that coincides with (and is tangential to) one 
of the shifted Fermi surfaces.

In Figs.~\ref{fig:FS_reconstruction}(b) and \ref{fig:FS_reconstruction}(e) we 
map $A(\bk,\omega)$ at the respective Fermi energies $\varepsilon_F$ for the 
same modulation vectors as in (a) and (b). The spectral functions both exhibit
a depletion of spectral weight near the antinodes and the residual spectral 
weight lies mostly along Fermi surface arcs, in broad agreement with 
angle-resolved photoemission spectroscopy (ARPES) 
experiments.\cite{Damascelli:2003} ARPES has shown that the pseudogap in 
underdoped cuprates is generically characterized by a gap on the antinodal 
portion of the Fermi surface.  
  
One widely studied scenario for the gap invokes the proximity of the underdoped 
cuprates to an antiferromagnetic insulating phase, and assumes either static or 
dynamical SDW correlations with a modulation wavevector near $(\pi,\pi)$. These 
kinds of scenarios generically lead to a spectral function with four hole-like 
Fermi surface pockets\cite{Atkinson:2012} centered near $(\pm\pi/2,\pm\pi/2)$,  
but such pockets have yet to be observed experimentally.  
In contrast, scenarios in which ${\bf q}^\ast$ nests antinodal Fermi surface 
sections do not generate nodal pockets; consistent with this the spectral functions in
Fig.~\ref{fig:FS_reconstruction}(b) and (e) have no backfolding around 
$(\pm \pi/2,\pm \pi/2)$.  
  
It is obvious from Figs.~\ref{fig:FS_reconstruction}(b) and 
\ref{fig:FS_reconstruction}(e)
that unidirectional MNO leads to an orthorhombic distortion of the Fermi 
surface.
Four-fold rotational symmetry is restored when the MNO is bidirectional, as 
illustrated by Figs.~\ref{fig:FS_reconstruction}(c) and 
\ref{fig:FS_reconstruction}(f). The spectral functions for bidirectional MNO 
are qualitatively similar to the unidirectional cases, with one notable 
exception: the reconstructed Fermi surface in 
Fig.~\ref{fig:FS_reconstruction}(c) has small electron-like pockets that close 
around $(\pi,0)$ and $(0,\pi)$. These pockets have low spectral weight, and 
therefore do not show up strongly in the spectral function. One of the pockets
is shown in the inset to Fig.~\ref{fig:FS_reconstruction}(c), where a portion 
of the reconstructed Fermi surface is plotted. We discuss these pockets further 
in Sec.~\ref{sec:discussionB}.
 
\section{Discussion}
\label{sec:discussion}
\subsection{Comparison to One-Band Models}
\label{sec:discussionA}
In this section we address the doping dependence of the modulation wavevector 
$\bq^\ast$. We have noted that for $t_{pp}=0.5$, axial MNO is preferred for 
$p>p_\mathrm{vH}$ and diagonal MNO is preferred for $p<p_\mathrm{vH}$ as was also
obtained by Holder and Metzner.\cite{Holder:2012ks} In their one-band model the 
nematic instabilities involve a spontaneous symmetry breaking between the $x$- 
and $y$-axis bond order parameters $\langle c_{i+x\,\sigma}^\dagger c_{i\,\sigma}
\rangle$ and $\langle c_{i+y\,\sigma}^\dagger c_{i\,\sigma}\rangle$, while in the 
three-band model, the instability involves a transfer of charge between
O$p_x$ and O$p_y$ orbitals.  

\begin{figure}
\includegraphics[width=\columnwidth]{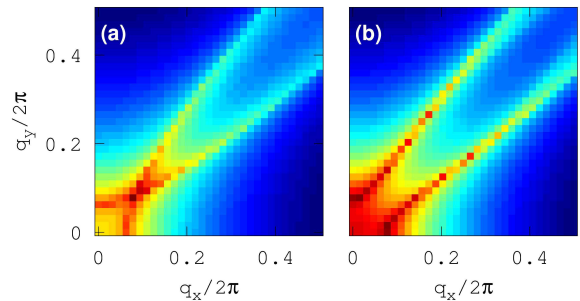}
\caption{Bare nematic susceptibility $\chi^0_\mathrm{1B}(\bq)$ for the 
conduction band of the three-band model.  The susceptibility is calculated from 
Eq.~\reff{eq:1B}, with $\epsilon_\bk$ set equal to the largest eigenvalue 
$E^3_\bk$ of the Hamiltonian ${\bf H}_0(\bk)$. Results are shown for hole 
dopings (a) $p=0.14 < p_\mathrm{vH}$ and (b) $p=0.20> p_\mathrm{vH}$, with 
$t_{pp}= 0.5$.}
\label{fig:chi01B}
\end{figure}
As emphasized in Ref.~\onlinecite{Holder:2012ks}, the nematic instability in the
one-band model is understood from the peak structure of the $d$-wave 
polarization function 
\begin{equation}
\chi^{0}_\mathrm{1B}(\bq)= -\frac{1}{N}\sum_{\bk}\frac{f(\epsilon_{\bk-\bq/2})-
f(\epsilon_{\bk+\bq/2})}{\epsilon_{\bk-\bq/2}-\epsilon_{\bk+\bq/2}} d_\bk^2
\label{eq:1B}
\end{equation}
where $\epsilon_\bk$ is the band dispersion and $d_\bk=\cos(k_x)-\cos(k_y)$ is a
$d$-wave form factor. $\chi^0_{1B}({\bf q})$ is illustrated in
Fig.~\ref{fig:chi01B} for electron densities below and above the van Hove
filling. For this figure, we have set $\epsilon_\bk =E^3_\bk$, where $E^3_\bk$ 
is the conduction band dispersion for the three-band model, obtained by 
diagonalizing ${\bf H}_0(\bk)$.

The ${\bf q}$-space arcs along which $\chi^0_\mathrm{1B}(\bq)$ is peaked in
Fig.~\ref{fig:chi01B} correspond to the Fermi surface nesting condition 
$E^3_{\bk} = E^3_{\bk+\bq}=\varepsilon_F$, with $\varepsilon_F$ the Fermi energy.
For hole-like Fermi surfaces, these arcs cross [Fig.~\ref{fig:chi01B}(a)] so 
that a single peak on the Brillouin zone diagonal emerges at the crossing point
$\bq^\ast$. For electron-like Fermi surfaces instead, the arcs do not cross
[Fig.~\ref{fig:chi01B}(b)] and two maxima lie on the $x$- and the $y$-axis, 
respectively. While this simple analysis explains the doping dependence of
$\bq^\ast$ in the one-band model, its applicability to the three-band
model is not at all obvious.

\begin{figure}
\includegraphics[width=\columnwidth]{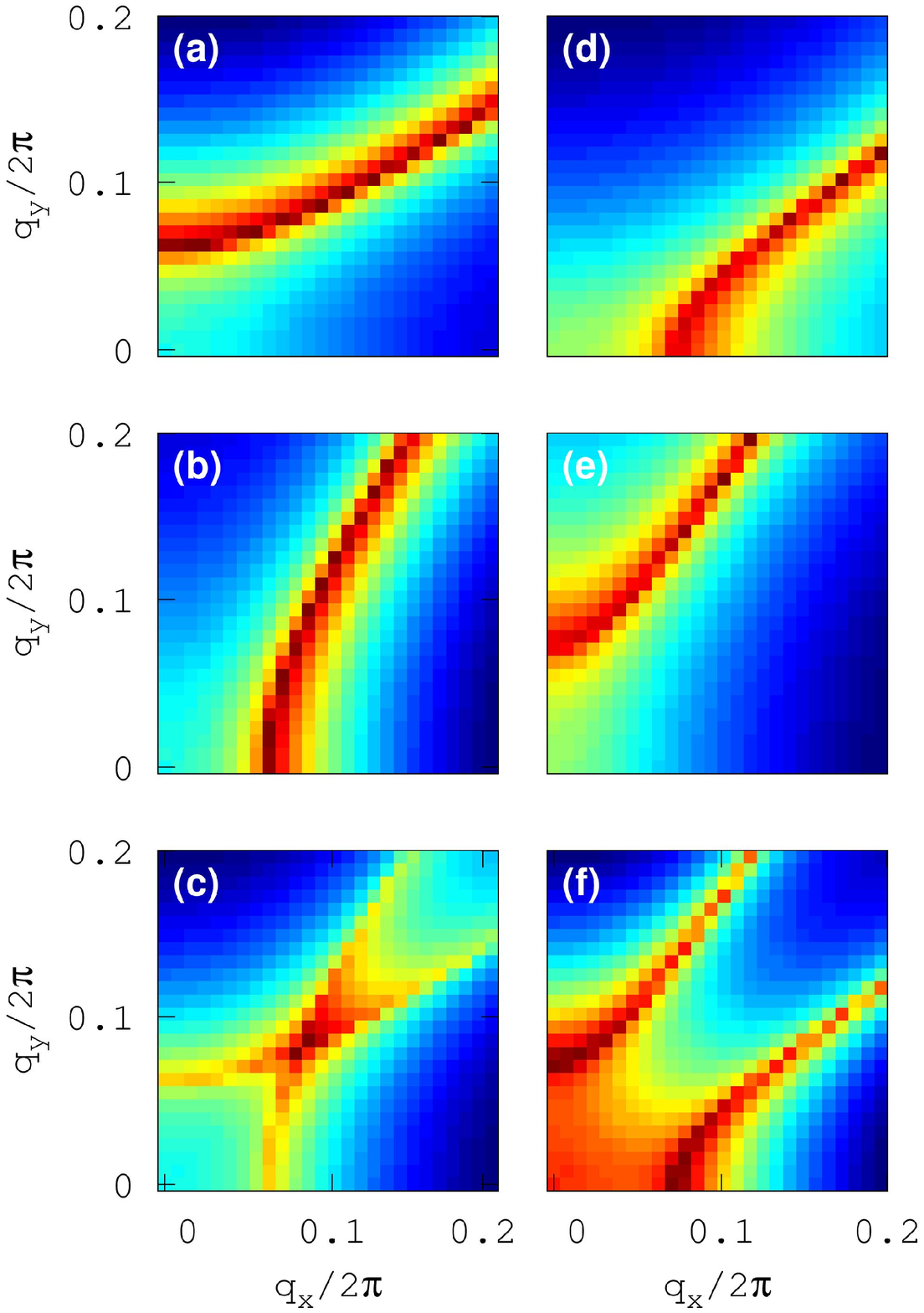}
\caption{Plots of (a),(d) $\chi^0_{xx}(\bq)$, (b),(e) $\chi^0_{yy}(\bq)$, and 
(c),(f) $\overline \chi^0(\bq)$ [Eq.~\reff{e:vpp_cond}] for $T=0.0005$ and 
fillings (a)-(c) $p=0.14$ and (d)-(f) $p=0.20$. }
\label{f:vpp_div_cond}
\end{figure}

The bare susceptibility in the three-band model is a $3\times 3$ matrix,
\begin{eqnarray}
\chi^0_{\theta\gamma}(\bq) &=& -\frac{1}{N}\sum_{\bk} \sum_{\mu,\nu=1}^3\left 
|S_{\gamma\nu}\left (\bk-\frac{\bq}{2}\right )
S_{\theta\mu}\left (\bk+\frac{\bq}{2}\right )\right|^2 \nonumber \\
&& \times 
\frac{f(E^\nu_{\bk-\bq/2})-f(E^\mu_{\bk+\bq/2})}
{E^\nu_{\bk-\bq/2}-E^\mu_{\bk+\bq/2}},
\label{eq:3B}
\end{eqnarray}
where ${\bf S}(\bk)$ is the unitary matrix defined by Eq.~\reff{eq:Sk}
that diagonalizes ${\bf H}_0(\bk)$.  Empirically, the most important
contributions to the nematic susceptibility [Eq.~\reff{eq:chiN}] are from 
$\chi_{xx}^0(\bq)$ and $\chi_{yy}^0(\bq)$, which are at least an order of 
magnitude larger than $\chi_{xy}^0(\bq)$ and $\chi_{yx}^0(\bq)$. These two 
dominant contributions are illustrated in Figs.~\ref{f:vpp_div_cond}(a)-(d) for 
fillings below and above the van Hove filling.

The most important contribution to $\chi^0_{\theta\gamma}(\bq)$ in
Eq.~\reff{eq:3B} comes from intraband transitions in the conduction band 
($\mu=\nu=3$). Focusing on this contribution alone, the main distinction 
between Eq.~\reff{eq:3B} and Eq.~\reff{eq:1B} is the weighting factor, which in 
the one-band case is $d_\bk^2$ and in the three-band case consists of a product 
of matrix elements of ${\bf S}(\bk)$. As is apparent from the plots of
$\chi_{xx}^0(\bq)$ and $\chi_{yy}^0(\bq)$ in Fig.~\ref{f:vpp_div_cond}, these 
matrix elements break the fourfold rotational symmetry of the underlying band 
structure.  Rotational symmetry is restored, however, for the bare nematic 
susceptibility, which is approximately determined by
$\chi^0_{N}(\bq)\approx \chi^0_{xx}(\bq)+ \chi^0_{yy}(\bq)$.

Similarly, the full nematic susceptibility has a fourfold rotational symmetry, 
although the structure is somewhat different from that of $\chi^0_{N}(\bq)$. To 
show this, following Bang {\it et al.},\cite{Bang:1993} we evaluate 
Eq.~\reff{e:chiladder} for a reduced basis set comprising the most important 
basis functions, $g^9_{\alpha\beta}$, $g^{10}_{\alpha\beta}$, and 
$g^{11}_{\alpha\beta}$.  To obtain a qualitative picture of the transition, we 
keep only the contribution of $V_{pp}$, which drives the charge instability,
and set $U_d=V_{pd}=U_p=0$. The interaction in the charge channel is thus 
reduced to
\begin{equation} 
{\bf \tilde V_\rho} =2{\bf \tilde V_D} - {\bf \tilde V_X}
=\left [ 
\begin{array}{ccc}
	0 & 0 & 0 \\	
	0 & 0 & \bar V_{pp} \\
	0 & \bar V_{pp} & 0
\end{array}
\right ],
\end{equation}
where $\bar V_{pp}=8V_{pp}\cos(q_x/2)\cos(q_y/2)$. The transition to
a charge ordered state occurs when the smallest eigenvalue of the matrix
${\bf D} = {\bf 1}+{\bf \tilde V}_\rho{\tilde\chibf}_0$, the
inverse of which enters into Eq.~\reff{e:gamma}, vanishes. For the reduced basis
and the simplified interaction, this matrix is
\begin{eqnarray}
  {\bf D} 
  &=&\left [
    \begin{array}{ccc}
      1 & 0 & 0 \\
      \bar V_{pp}\chi_{dy}^0 & 1  & \bar V_{pp} \chi_{yy}^0 \\
      \bar V_{pp}\chi_{dx}^0  & \bar V_{pp}\chi_{xx}^0  & 1 
    \end{array} 
    \right ].
\end{eqnarray}
It is straightforward to show that the smallest eigenvalue of ${\bf D}$ is 
$1-\overline{V}_{pp} \overline \chi^0(\bq)$, where 
\begin{equation}
\overline \chi^0(\bq) = 
  \sqrt{\chi^0_{xx}(\bq)\chi^0_{yy}(\bq)},
  \label{e:vpp_cond}
\end{equation} 
is the geometric average of $\chi^0_{xx}(\bq)$ and $\chi^0_{yy}(\bq)$. The 
quantity $\overline \chi^0(\bq)$ can be interpreted as the bare susceptibility 
for the charge mode associated with the leading nematic instability. As one can 
see from a comparison of Figs.~\ref{fig:chi01B}(a),(b) and 
Figs.~\ref{f:vpp_div_cond}(e),(f), $\overline \chi^0(\bq)$ and
$\chi^0_\mathrm{1B}(\bq)$ are qualitatively similar. In particular, 
$\overline \chi^0(\bq)$ has the same arc structure as in the one-band model: 
$\bq^\ast$ lies along the Brillouin zone diagonal when the arcs cross (hole-like
Fermi surface) and it lies on the Brillouin zone axes when the arcs do not cross
(electron-like Fermi surface).  The connection between the direction of 
$\bq^\ast$ and the Fermi surface topology thus appears to be a robust feature.

\subsection{Comparison to Experiments}
\label{sec:discussionB}
\begin{figure}
\includegraphics[width=\columnwidth]{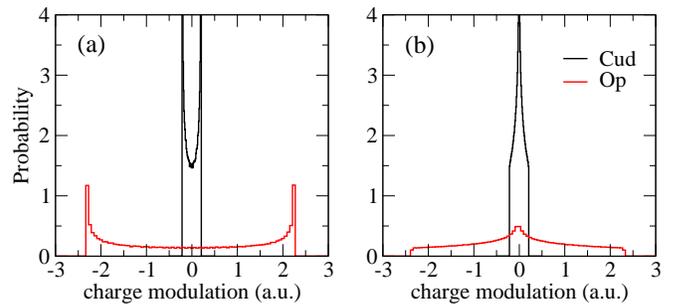}
  \caption{Normalized histograms of the charge modulations in
    Fig.~\ref{fig:CDWpattern} for $p=0.14$ (diagonal MNO) with (a)
    unidirectional and (b) bidirectional modulations.  Histograms are
    shown for both Cu$d$ and O$p$ orbitals.  The distributions for
    axial MNO are similar.  }
  \label{fig:charge_hist}
\end{figure}

Our calculations lead to an incommensurate modulation with a doping dependent 
$\bq^\ast$. Experimentally, there has been some debate about the value and the
doping dependence of the modulation wavevector. NMR studies of ortho-II YBCO 
(YBa$_2$Cu$_3$O$_{6.54}$) found that the planar Cu NMR line is split into two 
distinct peaks corresponding to Cu sites below empty (E) and filled (F) CuO 
chains (in ortho-II YBCO, every second chain has no oxygen), and that the NMR 
line corresponding to Cu-E sites develops a double-peak at the onset of charge 
order.\cite{Wu:2011ke} This was interpreted in terms of commensurate period-4 
order, with $q^\ast=0.25$ in units of $2\pi/a_0$, where $a_0$ is the lattice
constant. Subsequent x-ray scattering 
experiments\cite{Ghiringhelli:2012bw,Chang:2012vf} found incommensurate 
modulations with $q^\ast \approx 0.32$. This appears at first sight to be 
inconsistent with the NMR experiments because an incommensurate modulation 
should produce a continuous, rather than bimodal, distribution of Cu$d$ charge 
densities. However, it was recently pointed out\cite{Hayden:2013MM} that the 
distribution for an incommensurate charge modulation has a double-peak structure
provided the modulations are unidirectional. To illustrate this, we plot in
Fig.~\ref{fig:charge_hist} the normalized histograms of both the Cu$d$ and the 
O$p$ charge-density modulations shown in Figs.~\ref{fig:CDWpattern}(a)
and \ref{fig:CDWpattern}(c). For the unidirectional MNO,
Fig.~\ref{fig:charge_hist}(a), the distribution has the double-peak structure 
described above. The width of the distribution is larger for the O$p$ orbitals 
than for the Cu$d$ orbitals because nematic order involves primarily the oxygen 
sites. The double-peak structure is not present when the MNO is bidirectional,
Fig.~\ref{fig:charge_hist}(b), so consistency with NMR requires that ortho-II 
charge modulations be unidirectional. Recent x-ray experiments by Blackburn 
{\it et al.} did indeed observe that charge modulations in ortho-II 
YBCO are predominantly along the $b$-axis.\cite{Blackburn:2012wn_prl}

The same series of x-ray experiments finds modulation wavevectors along both the
$a$ and $b$ directions at higher dopings. An interesting question, yet to be 
addressed, is whether NMR is consistent with these experiments. In particular, 
NMR experiments should be able to determine, based on whether the lineshape 
resembles those in Figs.~\ref{fig:charge_hist}(a) or \ref{fig:charge_hist}(b),  
whether the two $q^\ast$ values correspond to spatially separated domains of 
unidirectional order or to bidirectional order.

The doping dependence of $q^\ast$ also needs to be resolved experimentally. The 
MNO described by our calculations originates from a Fermi surface instability 
and $q^\ast$ therefore has a strong doping dependence: for a hole-like Fermi 
surface, the modulation wavevector decreases as the hole density increases and 
vanishes at $p_\mathrm{vH}$. Ghiringhelli {\it et al.}\cite{Ghiringhelli:2012bw}
did not find any doping dependence to the modulation wavevector, but Blackburn 
{\it et al.}\cite{Blackburn:2012wn_prl} found a $\sim 10\%$ decrease in $q^\ast$ as 
the hole doping was increased from $0.104$ to $0.132$ (a 30\% change). Although 
the variation of $q^\ast$ with doping is slower than predicted by our 
calculations, the general trend is qualitatively consistent.  The doping
dependence of checkerboard modulations has also been explored by STM in other materials:
the modulation wavevector was found to be weakly doping dependent in 
$\mathrm{Bi_2Sr_2CaCu_2O_{8+\delta}}$ (Bi2212)\cite{Kohsaka:2008} and
in $\mathrm{Ca_{2-x}Na_xCuO_2Cl_2}$ (Ref.~\onlinecite{Hanaguri:2004}), but is
much more strongly doping dependent in $\mathrm{Bi_{2-y} Pb_y Sr_{2-z} La_z CuO_{6+x}}$
(Bi2201)\cite{Wise:2008}.  In Bi2201, the modulation wavevector decreased from $2\pi/4.5 a_0$
to $2\pi/6.2a_0$ as the doping was changed from  underdoped ($T_c=25$ K) to optimal ($T_c=35$ K).  These $q^\ast$ values are much smaller than what is found in YBCO, and are quantitatively close to what we have found, though the ordering is rotated by $45^\circ$.

As we have mentioned above, the two main discrepancies between our calculations 
and the experiments on YBCO are that for hole-like Fermi surfaces the calculated 
orientation of $q^\ast$ is diagonal rather than axial, and that the magnitude of
$q^\ast$ is too small by a factor of 2 or 3. The magnitude of $q^\ast$ in our 
calculations is set by the Fermi surface structure, and it is entirely possible 
that the discrepancy may be reconciled by tailoring the model specifically to 
ortho-II YBCO.  Such a model could include, for example, additional Fermi
surface sheets due to CuO chains, or to the fact that each unit cell contains 
two CuO$_2$ planes.  With the exception of ortho-II YBCO, the CuO chains seem 
unlikely to resolve the discrepancies because charge order is observed along 
both $a$ and $b$ axes in detwinned crystals, where the chains lie along the $b$ 
direction. The bilayer structure, however, may be important. Ghiringhelli {\it 
et al.} noted that the observed $q^\ast$ is generally too large to be due to 
nesting of a single CuO$_2$ Fermi surface, and proposed instead that $q^\ast$
connects segments of the bonding Fermi surface of the CuO$_2$ bilayer. To 
explore this, we have repeated our calculations for a single CuO$_2$ bilayer, 
and have found that (i) it is possible for the susceptibility to be larger on 
the bonding sheet than on the antibonding sheet but that (ii) we always obtain 
diagonal MNO for hole-like Fermi surfaces. As discussed in the previous section,
the trend for hole-like Fermi surfaces to have a diagonal modulation vector is a
remarkably robust feature of our calculation.

Finally, we return to the discussion of the Fermi surface in the modulated 
nematic phase. As we noted in Sec.~\ref{sec:FermiSurface}, the spectral 
intensity is suppressed in the antinodal region, and is therefore broadly 
consistent with photoemission experiments on underdoped cuprates. We also noted 
that for hole-like Fermi surfaces (corresponding to underdoped cuprates) the 
reconstructed Fermi surface has small electron-like pockets near $(\pi,0)$ and 
$(0,\pi)$. This is particularly interesting because magneto-oscillation
experiments,\cite{DoironLeyraud:2007bj,Sebastian:2012wh,Riggs:2011ii,Singleton:2010bz,Bangura:2008cg,Yelland:2008dh,Sebastian:2008hp,Sebastian:2012bl}
along with Hall measurements,\cite{LeBoeuf:2007gi} on YBCO have demonstrated the
existence of small electron-like Fermi surface pockets occupying $\sim 2\%$ of 
the Brillouin zone area. We make a few observations about a possible 
correspondence between the experiments and our calculations. First, while 
electron pockets should be observable in ARPES experiments, such experiments are
difficult to perform on YBCO. Second, a large number of ARPES experiments has 
been done on Bi-based cuprates, and these have not seen electron pockets. 
However, Bi-cuprates are highly inhomogeneous and this can mask spectral 
signatures of non-superconducting phases.\cite{Atkinson:2012} Furthermore, 
magneto-oscillation experiments are hampered in the Bi-cuprates by high levels 
of disorder, and have not provided independent confirmation of the Fermi surface
structure. It is thus possible that charge modulations are accompanied by 
well-defined antinodal electron pockets in YBCO but not in Bi-cuprates. Third, 
the structure and the existence of the predicted pockets in our calculations
depend on the size and the orientation of $\bq^\ast$. At present it is unclear 
whether a model that correctly predicts the orientation of $\bq^\ast$ will also 
generate an antinodal electron pocket.

\section{Conclusions}
We have calculated, within a weak-coupling diagrammatic perturbation theory, the
charge susceptibility matrix $\chi_{\alpha\beta}(\bq,\omega)$, from which we have
identified several nematic charge instabilities in the three-band Emery model.
Taking model parameters appropriate for cuprate superconductors, we find that 
there is a broad doping range over which the model is unstable towards a 
modulated nematic phase, characterized by a charge transfer between oxygen 
$p_x$ and $p_y$ orbitals. Such a phase has many features consistent with 
experiments. However, in the relevant doping range, the orientation of the 
modulation wavevector is rotated with respect to that measured in x-ray 
scattering experiments. At this point, it is not clear whether this discrepancy 
is the result of an oversimplification of the cuprate band structure by the 
three-band model, or whether it is the result of still unresolved physics that 
is missing from the model.

\section*{Acknowledgments}
We thank B.~M.~Andersen and M.-H. Julien for helpful conversations. W.A.A. acknowledges support 
by the Natural Sciences and Engineering Research Council (NSERC) of Canada. This
work was made possible by the facilities of the Shared Hierarchical Academic 
Research Computing Network and Compute/Calcul Canada. S.B. acknowledges a German
Academic Exchange Service (DAAD) Research Grant PKZ1072498 received in the year
2011. A.P.K. acknowledges support by the DFG through TRR 80.

\appendix
\section{Function Basis}
\label{s:fbasis}
The momentum dependent interaction is separable in the following basis:
\begin{eqnarray}
g^1_{\alpha\beta}&=&g^{12}_{\beta\alpha}=\delta_{\alpha d}\delta_{\beta x} \cos(k_x/2), \\
g^2_{\alpha\beta}&=&g^{13}_{\beta\alpha}=\delta_{\alpha d}\delta_{\beta x} \sin(k_x/2), \\
g^3_{\alpha\beta}&=&g^{14}_{\beta\alpha}=\delta_{\alpha d}\delta_{\beta y} \cos(k_y/2), \\
g^4_{\alpha\beta}&=&g^{15}_{\beta\alpha}=\delta_{\alpha d}\delta_{\beta y} \sin(k_y/2), \\
g^5_{\alpha\beta}&=&g^{16}_{\beta\alpha}=\delta_{\alpha x}\delta_{\beta y} \cos(k_x/2)\cos(k_y/2), \\
g^6_{\alpha\beta}&=&g^{17}_{\beta\alpha}=\delta_{\alpha x}\delta_{\beta y} \cos(k_x/2)\sin(k_y/2), \\
g^7_{\alpha\beta}&=&g^{18}_{\beta\alpha}=\delta_{\alpha x}\delta_{\beta y} \sin(k_x/2)\cos(k_y/2), \\
g^8_{\alpha\beta}&=&g^{19}_{\beta\alpha}=\delta_{\alpha x}\delta_{\beta y} \sin(k_x/2)\sin(k_y/2), \\
g^{9}_{\alpha\beta}&=&\delta_{\alpha d}\delta_{\beta d}, \\
g^{10}_{\alpha\beta}&=&\delta_{\alpha x}\delta_{\beta x}, \\
g^{11}_{\alpha\beta}&=&\delta_{\alpha y}\delta_{\beta y}.
\end{eqnarray}
For example, the $p$-$d$ density-density interaction 
$V_{xd}(\bq)=2V_{pd}\cos(q_x/2)$ is 
\begin{eqnarray}
	V_{xd}(\bk-\bk^\prime)
	&= &2V_{pd}\cos(k_x/2-k^\prime_x/2) \\
	&=& 2V_{pd}\{ \cos(k_x/2)\cos(k^\prime_x/2) \nonumber \\
	& & +\sin(k_x/2)\sin(k^\prime_x/2) \} \\
	&=& \sum_{i=12}^{13}g^i_{xd}(\bk)\tilde V_X^{ii}g^i_{xd}(\bk^\prime)
\end{eqnarray}
where
\begin{equation}
		\tilde V_X^{ij} = \left \{
	\begin{array}{ccc}
		 2V_{pd}\delta_{ij}, &  i\in\{1\dots4,12\dots15\} \\
		 2V_{pp}\delta_{ij}, & i\in\{5\dots8,16\dots19\} \\
		 U_d \delta_{ij}, & i=9 \\
		 U_p \delta_{ij}, & i=10,11
	\end{array} \right .
\end{equation} 
is the exchange interaction matrix in this basis. The direct interaction 
$\tilde V_D^{ij}(\bq)$ is zero everywhere except for $i,j\in \{9,10,11\}$. In 
this $3\times 3$ block
\begin{equation}
	\tilde V_D(\bq) \label{e:vcharge}
	= \left (
	\begin{array}{ccc}
       U_d &  2V_{pd}\text{c}_x(\bq) & 2V_{pd}\text{c}_y(\bq) \\ 
       2V_{pd}\text{c}_x(\bq) &  U_p  & 4V_{pp}\text{c}_x(\bq)\text{c}_y(\bq) \\ 
       2V_{pd}\text{c}_y(\bq) & 4V_{pp}\text{c}_x(\bq)\text{c}_y(\bq)	 & U_p 
       \end{array} 
       \right ) 
\end{equation} 
where $\text{c}_x(\bq)=\cos(q_x/2)$ and $\text{c}_y(\bq)=\cos(q_y/2)$.


\begin{thebibliography}{63}
\expandafter\ifx\csname natexlab\endcsname\relax\def\natexlab#1{#1}\fi
\expandafter\ifx\csname bibnamefont\endcsname\relax
  \def\bibnamefont#1{#1}\fi
\expandafter\ifx\csname bibfnamefont\endcsname\relax
  \def\bibfnamefont#1{#1}\fi
\expandafter\ifx\csname citenamefont\endcsname\relax
  \def\citenamefont#1{#1}\fi
\expandafter\ifx\csname url\endcsname\relax
  \def\url#1{\texttt{#1}}\fi
\expandafter\ifx\csname urlprefix\endcsname\relax\def\urlprefix{URL }\fi
\providecommand{\bibinfo}[2]{#2}
\providecommand{\eprint}[2][]{\url{#2}}

\bibitem[{\citenamefont{Tranquada et~al.}(1995)\citenamefont{Tranquada,
  Sternlieb, Axe, Nakamura, and Uchida}}]{Tranquada:1995}
\bibinfo{author}{\bibfnamefont{J.~M.} \bibnamefont{Tranquada}},
  \bibinfo{author}{\bibfnamefont{B.~J.} \bibnamefont{Sternlieb}},
  \bibinfo{author}{\bibfnamefont{J.~D.} \bibnamefont{Axe}},
  \bibinfo{author}{\bibfnamefont{Y.}~\bibnamefont{Nakamura}}, \bibnamefont{and}
  \bibinfo{author}{\bibfnamefont{S.}~\bibnamefont{Uchida}},
  \bibinfo{journal}{Nature} \textbf{\bibinfo{volume}{375}},
  \bibinfo{pages}{561} (\bibinfo{year}{1995}).

\bibitem[{\citenamefont{Abbamonte et~al.}(2005)\citenamefont{Abbamonte, Rusydi,
  Smadici, Gu, Sawatzky, and Feng}}]{Abbamonte:2005}
\bibinfo{author}{\bibfnamefont{P.}~\bibnamefont{Abbamonte}},
  \bibinfo{author}{\bibfnamefont{A.}~\bibnamefont{Rusydi}},
  \bibinfo{author}{\bibfnamefont{S.}~\bibnamefont{Smadici}},
  \bibinfo{author}{\bibfnamefont{G.~D.} \bibnamefont{Gu}},
  \bibinfo{author}{\bibfnamefont{G.~A.} \bibnamefont{Sawatzky}},
  \bibnamefont{and} \bibinfo{author}{\bibfnamefont{D.~L.} \bibnamefont{Feng}},
  \bibinfo{journal}{Nat. Phys.} \textbf{\bibinfo{volume}{1}},
  \bibinfo{pages}{155} (\bibinfo{year}{2005}).

\bibitem[{\citenamefont{Vojta}(2009)}]{Vojta:2009}
\bibinfo{author}{\bibfnamefont{M.}~\bibnamefont{Vojta}},
  \bibinfo{journal}{Advances in Physics} \textbf{\bibinfo{volume}{58}},
  \bibinfo{pages}{699} (\bibinfo{year}{2009}).

\bibitem[{\citenamefont{Tranquada et~al.}(2008)\citenamefont{Tranquada, Gu,
  H\"ucker, Jie, Kang, Klingeler, Li, Tristan, Wen, Xu, Xu, Zhou, and
  v.~Zimmermann}}]{Tranquada:2008}
\bibinfo{author}{\bibfnamefont{J.~M.} \bibnamefont{Tranquada}},
  \bibinfo{author}{\bibfnamefont{G.~D.} \bibnamefont{Gu}},
  \bibinfo{author}{\bibfnamefont{M.}~\bibnamefont{H\"ucker}},
  \bibinfo{author}{\bibfnamefont{Q.}~\bibnamefont{Jie}},
  \bibinfo{author}{\bibfnamefont{H.-J.} \bibnamefont{Kang}},
  \bibinfo{author}{\bibfnamefont{R.}~\bibnamefont{Klingeler}},
  \bibinfo{author}{\bibfnamefont{Q.}~\bibnamefont{Li}},
  \bibinfo{author}{\bibfnamefont{N.}~\bibnamefont{Tristan}},
  \bibinfo{author}{\bibfnamefont{J.~S.} \bibnamefont{Wen}},
  \bibinfo{author}{\bibfnamefont{G.~Y.} \bibnamefont{Xu}},
  \bibinfo{author}{\bibfnamefont{Z.~J.} \bibnamefont{Xu}},
  \bibinfo{author}{\bibfnamefont{J.}~\bibnamefont{Zhou}}, \bibnamefont{and}
  \bibinfo{author}{\bibfnamefont{M.}~\bibnamefont{v.~Zimmermann}},
  \bibinfo{journal}{Phys. Rev. B} \textbf{\bibinfo{volume}{78}},
  \bibinfo{pages}{174529} (\bibinfo{year}{2008}).

\bibitem[{\citenamefont{H\"ucker et~al.}(2013)\citenamefont{H\"ucker,
  v.~Zimmermann, Xu, Wen, Gu, and Tranquada}}]{Hucker:2012tv}
\bibinfo{author}{\bibfnamefont{M.}~\bibnamefont{H\"ucker}},
  \bibinfo{author}{\bibfnamefont{M.}~\bibnamefont{v.~Zimmermann}},
  \bibinfo{author}{\bibfnamefont{Z.~J.} \bibnamefont{Xu}},
  \bibinfo{author}{\bibfnamefont{J.~S.} \bibnamefont{Wen}},
  \bibinfo{author}{\bibfnamefont{G.~D.} \bibnamefont{Gu}}, \bibnamefont{and}
  \bibinfo{author}{\bibfnamefont{J.~M.} \bibnamefont{Tranquada}},
  \bibinfo{journal}{Phys. Rev. B} \textbf{\bibinfo{volume}{87}},
  \bibinfo{pages}{014501} (\bibinfo{year}{2013}).

\bibitem[{\citenamefont{Doiron-Leyraud
  et~al.}(2007)\citenamefont{Doiron-Leyraud, Proust, LeBoeuf, Levallois,
  Bonnemaison, Liang, Bonn, Hardy, and Taillefer}}]{DoironLeyraud:2007bj}
\bibinfo{author}{\bibfnamefont{N.}~\bibnamefont{Doiron-Leyraud}},
  \bibinfo{author}{\bibfnamefont{C.}~\bibnamefont{Proust}},
  \bibinfo{author}{\bibfnamefont{D.}~\bibnamefont{LeBoeuf}},
  \bibinfo{author}{\bibfnamefont{J.}~\bibnamefont{Levallois}},
  \bibinfo{author}{\bibfnamefont{J.-B.} \bibnamefont{Bonnemaison}},
  \bibinfo{author}{\bibfnamefont{R.}~\bibnamefont{Liang}},
  \bibinfo{author}{\bibfnamefont{D.~A.} \bibnamefont{Bonn}},
  \bibinfo{author}{\bibfnamefont{W.~N.} \bibnamefont{Hardy}}, \bibnamefont{and}
  \bibinfo{author}{\bibfnamefont{L.}~\bibnamefont{Taillefer}},
  \bibinfo{journal}{Nature} \textbf{\bibinfo{volume}{447}},
  \bibinfo{pages}{565} (\bibinfo{year}{2007}).

\bibitem[{\citenamefont{Sebastian
  et~al.}(2012{\natexlab{a}})\citenamefont{Sebastian, Harrison, and
  Lonzarich}}]{Sebastian:2012wh}
\bibinfo{author}{\bibfnamefont{S.~E.} \bibnamefont{Sebastian}},
  \bibinfo{author}{\bibfnamefont{N.}~\bibnamefont{Harrison}}, \bibnamefont{and}
  \bibinfo{author}{\bibfnamefont{G.}~\bibnamefont{Lonzarich}},
  \bibinfo{journal}{Rep. Prog. Phys.} \textbf{\bibinfo{volume}{75}},
  \bibinfo{pages}{102501} (\bibinfo{year}{2012}{\natexlab{a}}).

\bibitem[{\citenamefont{LeBoeuf et~al.}(2007)\citenamefont{LeBoeuf,
  Doiron-Leyraud, Levallois, Daou, Bonnemaison, Hussey, Balicas, Ramshaw,
  Liang, Bonn, Hardy, Adachi, Proust, and Taillefer}}]{LeBoeuf:2007gi}
\bibinfo{author}{\bibfnamefont{D.}~\bibnamefont{LeBoeuf}},
  \bibinfo{author}{\bibfnamefont{N.}~\bibnamefont{Doiron-Leyraud}},
  \bibinfo{author}{\bibfnamefont{J.}~\bibnamefont{Levallois}},
  \bibinfo{author}{\bibfnamefont{R.}~\bibnamefont{Daou}},
  \bibinfo{author}{\bibfnamefont{J.~B.} \bibnamefont{Bonnemaison}},
  \bibinfo{author}{\bibfnamefont{N.~E.} \bibnamefont{Hussey}},
  \bibinfo{author}{\bibfnamefont{L.}~\bibnamefont{Balicas}},
  \bibinfo{author}{\bibfnamefont{B.~J.} \bibnamefont{Ramshaw}},
  \bibinfo{author}{\bibfnamefont{R.}~\bibnamefont{Liang}},
  \bibinfo{author}{\bibfnamefont{D.~A.} \bibnamefont{Bonn}},
  \bibinfo{author}{\bibfnamefont{W.~N.} \bibnamefont{Hardy}},
  \bibinfo{author}{\bibfnamefont{S.}~\bibnamefont{Adachi}},
  \bibinfo{author}{\bibfnamefont{C.}~\bibnamefont{Proust}}, \bibnamefont{and}
  \bibinfo{author}{\bibfnamefont{L.}~\bibnamefont{Taillefer}},
  \bibinfo{journal}{Nature} \textbf{\bibinfo{volume}{450}},
  \bibinfo{pages}{533} (\bibinfo{year}{2007}).

\bibitem[{\citenamefont{Chakravarty}(2011)}]{Chakravarty:2011}
\bibinfo{author}{\bibfnamefont{S.}~\bibnamefont{Chakravarty}},
  \bibinfo{journal}{Rep. Prog. Phys.} \textbf{\bibinfo{volume}{74}},
  \bibinfo{pages}{022501} (\bibinfo{year}{2011}).

\bibitem[{\citenamefont{Daou et~al.}(2010)\citenamefont{Daou, Chang, LeBoeuf,
  Cyr-Choini{\`e}re, Lalibert{\'e}, Doiron-Leyraud, Ramshaw, Liang, Bonn,
  Hardy, and Taillefer}}]{Daou:2010bo}
\bibinfo{author}{\bibfnamefont{R.}~\bibnamefont{Daou}},
  \bibinfo{author}{\bibfnamefont{J.}~\bibnamefont{Chang}},
  \bibinfo{author}{\bibfnamefont{D.}~\bibnamefont{LeBoeuf}},
  \bibinfo{author}{\bibfnamefont{O.}~\bibnamefont{Cyr-Choini{\`e}re}},
  \bibinfo{author}{\bibfnamefont{F.}~\bibnamefont{Lalibert{\'e}}},
  \bibinfo{author}{\bibfnamefont{N.}~\bibnamefont{Doiron-Leyraud}},
  \bibinfo{author}{\bibfnamefont{B.~J.} \bibnamefont{Ramshaw}},
  \bibinfo{author}{\bibfnamefont{R.}~\bibnamefont{Liang}},
  \bibinfo{author}{\bibfnamefont{D.~A.} \bibnamefont{Bonn}},
  \bibinfo{author}{\bibfnamefont{W.~N.} \bibnamefont{Hardy}}, \bibnamefont{and}
  \bibinfo{author}{\bibfnamefont{L.}~\bibnamefont{Taillefer}},
  \bibinfo{journal}{Nature} \textbf{\bibinfo{volume}{463}},
  \bibinfo{pages}{519} (\bibinfo{year}{2010}).

\bibitem[{\citenamefont{Wu et~al.}(2011)\citenamefont{Wu, Mayaffre, Kr{\"a}mer,
  Horvati{\'c}, Berthier, Hardy, Liang, Bonn, and Julien}}]{Wu:2011ke}
\bibinfo{author}{\bibfnamefont{T.}~\bibnamefont{Wu}},
  \bibinfo{author}{\bibfnamefont{H.}~\bibnamefont{Mayaffre}},
  \bibinfo{author}{\bibfnamefont{S.}~\bibnamefont{Kr{\"a}mer}},
  \bibinfo{author}{\bibfnamefont{M.}~\bibnamefont{Horvati{\'c}}},
  \bibinfo{author}{\bibfnamefont{C.}~\bibnamefont{Berthier}},
  \bibinfo{author}{\bibfnamefont{W.~N.} \bibnamefont{Hardy}},
  \bibinfo{author}{\bibfnamefont{R.}~\bibnamefont{Liang}},
  \bibinfo{author}{\bibfnamefont{D.~A.} \bibnamefont{Bonn}}, \bibnamefont{and}
  \bibinfo{author}{\bibfnamefont{M.-H.} \bibnamefont{Julien}},
  \bibinfo{journal}{Nature} \textbf{\bibinfo{volume}{477}},
  \bibinfo{pages}{191} (\bibinfo{year}{2011}).

\bibitem[{\citenamefont{Hinkov et~al.}(2007)\citenamefont{Hinkov, Bourges,
  Pailh{\`e}s, Sidis, Ivanov, Frost, Perring, Lin, Chen, and
  Keimer}}]{Hinkov:2007gq}
\bibinfo{author}{\bibfnamefont{V.}~\bibnamefont{Hinkov}},
  \bibinfo{author}{\bibfnamefont{P.}~\bibnamefont{Bourges}},
  \bibinfo{author}{\bibfnamefont{S.}~\bibnamefont{Pailh{\`e}s}},
  \bibinfo{author}{\bibfnamefont{Y.}~\bibnamefont{Sidis}},
  \bibinfo{author}{\bibfnamefont{A.}~\bibnamefont{Ivanov}},
  \bibinfo{author}{\bibfnamefont{C.~D.} \bibnamefont{Frost}},
  \bibinfo{author}{\bibfnamefont{T.~G.} \bibnamefont{Perring}},
  \bibinfo{author}{\bibfnamefont{C.~T.} \bibnamefont{Lin}},
  \bibinfo{author}{\bibfnamefont{D.~P.} \bibnamefont{Chen}}, \bibnamefont{and}
  \bibinfo{author}{\bibfnamefont{B.}~\bibnamefont{Keimer}},
  \bibinfo{journal}{Nat. Phys.} \textbf{\bibinfo{volume}{3}},
  \bibinfo{pages}{780} (\bibinfo{year}{2007}).

\bibitem[{\citenamefont{Hinkov et~al.}(2008)\citenamefont{Hinkov, Haug,
  Fauqu\'{e}, Bourges, Sidis, Ivanov, Bernhard, Lin, and
  Keimer}}]{Hinkov:2008sci}
\bibinfo{author}{\bibfnamefont{V.}~\bibnamefont{Hinkov}},
  \bibinfo{author}{\bibfnamefont{D.}~\bibnamefont{Haug}},
  \bibinfo{author}{\bibfnamefont{B.}~\bibnamefont{Fauqu\'{e}}},
  \bibinfo{author}{\bibfnamefont{P.}~\bibnamefont{Bourges}},
  \bibinfo{author}{\bibfnamefont{Y.}~\bibnamefont{Sidis}},
  \bibinfo{author}{\bibfnamefont{A.}~\bibnamefont{Ivanov}},
  \bibinfo{author}{\bibfnamefont{C.}~\bibnamefont{Bernhard}},
  \bibinfo{author}{\bibfnamefont{C.~T.} \bibnamefont{Lin}}, \bibnamefont{and}
  \bibinfo{author}{\bibfnamefont{B.}~\bibnamefont{Keimer}},
  \bibinfo{journal}{Science} \textbf{\bibinfo{volume}{319}},
  \bibinfo{pages}{597} (\bibinfo{year}{2008}).

\bibitem[{\citenamefont{Haug et~al.}(2010)\citenamefont{Haug, Hinkov, Sidis,
  Bourges, Christensen, Ivanov, Keller, Lin, and Keimer}}]{Haug:2010ei}
\bibinfo{author}{\bibfnamefont{D.}~\bibnamefont{Haug}},
  \bibinfo{author}{\bibfnamefont{V.}~\bibnamefont{Hinkov}},
  \bibinfo{author}{\bibfnamefont{Y.}~\bibnamefont{Sidis}},
  \bibinfo{author}{\bibfnamefont{P.}~\bibnamefont{Bourges}},
  \bibinfo{author}{\bibfnamefont{N.~B.} \bibnamefont{Christensen}},
  \bibinfo{author}{\bibfnamefont{A.}~\bibnamefont{Ivanov}},
  \bibinfo{author}{\bibfnamefont{T.}~\bibnamefont{Keller}},
  \bibinfo{author}{\bibfnamefont{C.~T.} \bibnamefont{Lin}}, \bibnamefont{and}
  \bibinfo{author}{\bibfnamefont{B.}~\bibnamefont{Keimer}},
  \bibinfo{journal}{New J. Phys.} \textbf{\bibinfo{volume}{12}},
  \bibinfo{pages}{105006} (\bibinfo{year}{2010}).

\bibitem[{\citenamefont{Ghiringhelli et~al.}(2012)\citenamefont{Ghiringhelli,
  Le~Tacon, Minola, Blanco-Canosa, Mazzoli, Brookes, De~Luca, Frano, Hawthorn,
  He, Loew, Sala, Peets, Salluzzo, Schierle, Sutarto, Sawatzky, Weschke,
  Keimer, and Braicovich}}]{Ghiringhelli:2012bw}
\bibinfo{author}{\bibfnamefont{G.}~\bibnamefont{Ghiringhelli}},
  \bibinfo{author}{\bibfnamefont{M.}~\bibnamefont{Le~Tacon}},
  \bibinfo{author}{\bibfnamefont{M.}~\bibnamefont{Minola}},
  \bibinfo{author}{\bibfnamefont{S.}~\bibnamefont{Blanco-Canosa}},
  \bibinfo{author}{\bibfnamefont{C.}~\bibnamefont{Mazzoli}},
  \bibinfo{author}{\bibfnamefont{N.~B.} \bibnamefont{Brookes}},
  \bibinfo{author}{\bibfnamefont{G.~M.} \bibnamefont{De~Luca}},
  \bibinfo{author}{\bibfnamefont{A.}~\bibnamefont{Frano}},
  \bibinfo{author}{\bibfnamefont{D.~G.} \bibnamefont{Hawthorn}},
  \bibinfo{author}{\bibfnamefont{F.}~\bibnamefont{He}},
  \bibinfo{author}{\bibfnamefont{T.}~\bibnamefont{Loew}},
  \bibinfo{author}{\bibfnamefont{M.~M.} \bibnamefont{Sala}},
  \bibinfo{author}{\bibfnamefont{D.~C.} \bibnamefont{Peets}},
  \bibinfo{author}{\bibfnamefont{M.}~\bibnamefont{Salluzzo}},
  \bibinfo{author}{\bibfnamefont{E.}~\bibnamefont{Schierle}},
  \bibinfo{author}{\bibfnamefont{R.}~\bibnamefont{Sutarto}},
  \bibinfo{author}{\bibfnamefont{G.~A.} \bibnamefont{Sawatzky}},
  \bibinfo{author}{\bibfnamefont{E.}~\bibnamefont{Weschke}},
  \bibinfo{author}{\bibfnamefont{B.}~\bibnamefont{Keimer}}, \bibnamefont{and}
  \bibinfo{author}{\bibfnamefont{L.}~\bibnamefont{Braicovich}},
  \bibinfo{journal}{Science} \textbf{\bibinfo{volume}{337}},
  \bibinfo{pages}{821} (\bibinfo{year}{2012}).

\bibitem[{\citenamefont{Chang et~al.}(2012)\citenamefont{Chang, Blackburn,
  Holmes, Christensen, Larsen, Mesot, Liang, Bonn, Hardy, Watenphul,
  Zimmermann, Forgan, and Hayden}}]{Chang:2012vf}
\bibinfo{author}{\bibfnamefont{J.}~\bibnamefont{Chang}},
  \bibinfo{author}{\bibfnamefont{E.}~\bibnamefont{Blackburn}},
  \bibinfo{author}{\bibfnamefont{A.~T.} \bibnamefont{Holmes}},
  \bibinfo{author}{\bibfnamefont{N.~B.} \bibnamefont{Christensen}},
  \bibinfo{author}{\bibfnamefont{J.}~\bibnamefont{Larsen}},
  \bibinfo{author}{\bibfnamefont{J.}~\bibnamefont{Mesot}},
  \bibinfo{author}{\bibfnamefont{R.}~\bibnamefont{Liang}},
  \bibinfo{author}{\bibfnamefont{D.~A.} \bibnamefont{Bonn}},
  \bibinfo{author}{\bibfnamefont{W.~N.} \bibnamefont{Hardy}},
  \bibinfo{author}{\bibfnamefont{A.}~\bibnamefont{Watenphul}},
  \bibinfo{author}{\bibfnamefont{M.~v.} \bibnamefont{Zimmermann}},
  \bibinfo{author}{\bibfnamefont{E.~M.} \bibnamefont{Forgan}},
  \bibnamefont{and} \bibinfo{author}{\bibfnamefont{S.~M.}
  \bibnamefont{Hayden}}, \bibinfo{journal}{Nat. Phys.}
  \textbf{\bibinfo{volume}{8}}, \bibinfo{pages}{871} (\bibinfo{year}{2012}).

\bibitem[{\citenamefont{Blackburn et~al.}(2013)\citenamefont{Blackburn, Chang,
  H\"ucker, Holmes, Christensen, Liang, Bonn, Hardy, R\"utt, Gutowski,
  Zimmermann, Forgan, and Hayden}}]{Blackburn:2012wn_prl}
\bibinfo{author}{\bibfnamefont{E.}~\bibnamefont{Blackburn}},
  \bibinfo{author}{\bibfnamefont{J.}~\bibnamefont{Chang}},
  \bibinfo{author}{\bibfnamefont{M.}~\bibnamefont{H\"ucker}},
  \bibinfo{author}{\bibfnamefont{A.~T.} \bibnamefont{Holmes}},
  \bibinfo{author}{\bibfnamefont{N.~B.} \bibnamefont{Christensen}},
  \bibinfo{author}{\bibfnamefont{R.}~\bibnamefont{Liang}},
  \bibinfo{author}{\bibfnamefont{D.~A.} \bibnamefont{Bonn}},
  \bibinfo{author}{\bibfnamefont{W.~N.} \bibnamefont{Hardy}},
  \bibinfo{author}{\bibfnamefont{U.}~\bibnamefont{R\"utt}},
  \bibinfo{author}{\bibfnamefont{O.}~\bibnamefont{Gutowski}},
  \bibinfo{author}{\bibfnamefont{M.~v.} \bibnamefont{Zimmermann}},
  \bibinfo{author}{\bibfnamefont{E.~M.} \bibnamefont{Forgan}},
  \bibnamefont{and} \bibinfo{author}{\bibfnamefont{S.~M.}
  \bibnamefont{Hayden}}, \bibinfo{journal}{Phys. Rev. Lett.}
  \textbf{\bibinfo{volume}{110}}, \bibinfo{pages}{137004}
  (\bibinfo{year}{2013}).

\bibitem[{\citenamefont{LeBoeuf et~al.}(2013)\citenamefont{LeBoeuf, Kr{\"a}mer,
  Hardy, Liang, Bonn, and Proust}}]{LeBoeuf:2012up_nphys}
\bibinfo{author}{\bibfnamefont{D.}~\bibnamefont{LeBoeuf}},
  \bibinfo{author}{\bibfnamefont{S.}~\bibnamefont{Kr{\"a}mer}},
  \bibinfo{author}{\bibfnamefont{W.~N.} \bibnamefont{Hardy}},
  \bibinfo{author}{\bibfnamefont{R.}~\bibnamefont{Liang}},
  \bibinfo{author}{\bibfnamefont{D.~A.} \bibnamefont{Bonn}}, \bibnamefont{and}
  \bibinfo{author}{\bibfnamefont{C.}~\bibnamefont{Proust}},
  \bibinfo{journal}{Nat. Phys.} \textbf{\bibinfo{volume}{9}},
  \bibinfo{pages}{79} (\bibinfo{year}{2013}).

\bibitem[{\citenamefont{Blanco-Canosa et~al.}(2013)\citenamefont{Blanco-Canosa,
  Frano, Loew, Lu, Porras, Ghiringhelli, Minola, Mazzoli, Braicovich, Schierle,
  Weschke, Le~Tacon, and Keimer}}]{Blanco-canosa:2013}
\bibinfo{author}{\bibfnamefont{S.}~\bibnamefont{Blanco-Canosa}},
  \bibinfo{author}{\bibfnamefont{A.}~\bibnamefont{Frano}},
  \bibinfo{author}{\bibfnamefont{T.}~\bibnamefont{Loew}},
  \bibinfo{author}{\bibfnamefont{Y.}~\bibnamefont{Lu}},
  \bibinfo{author}{\bibfnamefont{J.}~\bibnamefont{Porras}},
  \bibinfo{author}{\bibfnamefont{G.}~\bibnamefont{Ghiringhelli}},
  \bibinfo{author}{\bibfnamefont{M.}~\bibnamefont{Minola}},
  \bibinfo{author}{\bibfnamefont{C.}~\bibnamefont{Mazzoli}},
  \bibinfo{author}{\bibfnamefont{L.}~\bibnamefont{Braicovich}},
  \bibinfo{author}{\bibfnamefont{E.}~\bibnamefont{Schierle}},
  \bibinfo{author}{\bibfnamefont{E.}~\bibnamefont{Weschke}},
  \bibinfo{author}{\bibfnamefont{M.}~\bibnamefont{Le~Tacon}}, \bibnamefont{and}
  \bibinfo{author}{\bibfnamefont{B.}~\bibnamefont{Keimer}},
  \bibinfo{journal}{Phys. Rev. Lett.} \textbf{\bibinfo{volume}{110}},
  \bibinfo{pages}{187001} (\bibinfo{year}{2013}).

\bibitem[{\citenamefont{Wu et~al.}(2013)\citenamefont{Wu, Mayaffre, Kr\"{a}mer,
  Horvati\'{c}, Berthier, Kuhns, Reyes, Liang, Hardy, Bonn, and
  Julien}}]{Wu:2013}
\bibinfo{author}{\bibfnamefont{T.}~\bibnamefont{Wu}},
  \bibinfo{author}{\bibfnamefont{H.}~\bibnamefont{Mayaffre}},
  \bibinfo{author}{\bibfnamefont{S.}~\bibnamefont{Kr\"{a}mer}},
  \bibinfo{author}{\bibfnamefont{M.}~\bibnamefont{Horvati\'{c}}},
  \bibinfo{author}{\bibfnamefont{C.}~\bibnamefont{Berthier}},
  \bibinfo{author}{\bibfnamefont{P.~L.} \bibnamefont{Kuhns}},
  \bibinfo{author}{\bibfnamefont{A.~P.} \bibnamefont{Reyes}},
  \bibinfo{author}{\bibfnamefont{R.}~\bibnamefont{Liang}},
  \bibinfo{author}{\bibfnamefont{W.~N.} \bibnamefont{Hardy}},
  \bibinfo{author}{\bibfnamefont{D.~A.} \bibnamefont{Bonn}}, \bibnamefont{and}
  \bibinfo{author}{\bibfnamefont{M.-H.} \bibnamefont{Julien}},
  \bibinfo{journal}{Nat. Commun.} \textbf{\bibinfo{volume}{4}}
  (\bibinfo{year}{2013}).

\bibitem[{\citenamefont{Kondo et~al.}(2009)\citenamefont{Kondo, Khasanov,
  Takeuchi, Schmalian, and Kaminski}}]{Kondo:2009}
\bibinfo{author}{\bibfnamefont{T.}~\bibnamefont{Kondo}},
  \bibinfo{author}{\bibfnamefont{R.}~\bibnamefont{Khasanov}},
  \bibinfo{author}{\bibfnamefont{T.}~\bibnamefont{Takeuchi}},
  \bibinfo{author}{\bibfnamefont{J.}~\bibnamefont{Schmalian}},
  \bibnamefont{and} \bibinfo{author}{\bibfnamefont{A.}~\bibnamefont{Kaminski}},
  \bibinfo{journal}{Nature (London)} \textbf{\bibinfo{volume}{457}},
  \bibinfo{pages}{296} (\bibinfo{year}{2009}).

\bibitem[{\citenamefont{Hashimoto et~al.}(2010)\citenamefont{Hashimoto, He,
  Tanaka, Testaud, Meevasana, Moore, Lu, Yao, Yoshida, Eisaki, Devereaux,
  Hussain, and Shen}}]{Hashimoto:2010nphys}
\bibinfo{author}{\bibfnamefont{M.}~\bibnamefont{Hashimoto}},
  \bibinfo{author}{\bibfnamefont{R.-H.} \bibnamefont{He}},
  \bibinfo{author}{\bibfnamefont{K.}~\bibnamefont{Tanaka}},
  \bibinfo{author}{\bibfnamefont{J.-P.} \bibnamefont{Testaud}},
  \bibinfo{author}{\bibfnamefont{W.}~\bibnamefont{Meevasana}},
  \bibinfo{author}{\bibfnamefont{R.~G.} \bibnamefont{Moore}},
  \bibinfo{author}{\bibfnamefont{D.}~\bibnamefont{Lu}},
  \bibinfo{author}{\bibfnamefont{H.}~\bibnamefont{Yao}},
  \bibinfo{author}{\bibfnamefont{Y.}~\bibnamefont{Yoshida}},
  \bibinfo{author}{\bibfnamefont{H.}~\bibnamefont{Eisaki}},
  \bibinfo{author}{\bibfnamefont{T.~P.} \bibnamefont{Devereaux}},
  \bibinfo{author}{\bibfnamefont{Z.}~\bibnamefont{Hussain}}, \bibnamefont{and}
  \bibinfo{author}{\bibfnamefont{Z.-X.} \bibnamefont{Shen}},
  \bibinfo{journal}{Nat. Phys.} \textbf{\bibinfo{volume}{6}},
  \bibinfo{pages}{414} (\bibinfo{year}{2010}).

\bibitem[{\citenamefont{He et~al.}(2011)\citenamefont{He, Hashimoto,
  Karapetyan, Koralek, Hinton, Testaud, Nathan, Yoshida, Yao, Tanaka,
  Meevasana, Moore, Lu, Mo, Ishikado, Eisaki, Hussain, Devereaux, Kivelson,
  Orenstein, Kapitulnik, and Shen}}]{He:2011}
\bibinfo{author}{\bibfnamefont{R.-H.} \bibnamefont{He}},
  \bibinfo{author}{\bibfnamefont{M.}~\bibnamefont{Hashimoto}},
  \bibinfo{author}{\bibfnamefont{H.}~\bibnamefont{Karapetyan}},
  \bibinfo{author}{\bibfnamefont{J.~D.} \bibnamefont{Koralek}},
  \bibinfo{author}{\bibfnamefont{J.~P.} \bibnamefont{Hinton}},
  \bibinfo{author}{\bibfnamefont{J.~P.} \bibnamefont{Testaud}},
  \bibinfo{author}{\bibfnamefont{V.}~\bibnamefont{Nathan}},
  \bibinfo{author}{\bibfnamefont{Y.}~\bibnamefont{Yoshida}},
  \bibinfo{author}{\bibfnamefont{H.}~\bibnamefont{Yao}},
  \bibinfo{author}{\bibfnamefont{K.}~\bibnamefont{Tanaka}},
  \bibinfo{author}{\bibfnamefont{W.}~\bibnamefont{Meevasana}},
  \bibinfo{author}{\bibfnamefont{R.~G.} \bibnamefont{Moore}},
  \bibinfo{author}{\bibfnamefont{D.~H.} \bibnamefont{Lu}},
  \bibinfo{author}{\bibfnamefont{S.-K.} \bibnamefont{Mo}},
  \bibinfo{author}{\bibfnamefont{M.}~\bibnamefont{Ishikado}},
  \bibinfo{author}{\bibfnamefont{H.}~\bibnamefont{Eisaki}},
  \bibinfo{author}{\bibfnamefont{Z.}~\bibnamefont{Hussain}},
  \bibinfo{author}{\bibfnamefont{T.~P.} \bibnamefont{Devereaux}},
  \bibinfo{author}{\bibfnamefont{S.~A.} \bibnamefont{Kivelson}},
  \bibinfo{author}{\bibfnamefont{J.}~\bibnamefont{Orenstein}},
  \bibinfo{author}{\bibfnamefont{A.}~\bibnamefont{Kapitulnik}},
  \bibnamefont{and} \bibinfo{author}{\bibfnamefont{Z.-X.} \bibnamefont{Shen}},
  \bibinfo{journal}{Science} \textbf{\bibinfo{volume}{331}},
  \bibinfo{pages}{1579} (\bibinfo{year}{2011}).

\bibitem[{\citenamefont{Ideta et~al.}(2012)\citenamefont{Ideta, Yoshida,
  Fujimori, Anzai, Fujita, Ino, Arita, Namatame, Taniguchi, Shen, Takashima,
  Kojima, and Uchida}}]{Ideta:2012}
\bibinfo{author}{\bibfnamefont{S.-I.} \bibnamefont{Ideta}},
  \bibinfo{author}{\bibfnamefont{T.}~\bibnamefont{Yoshida}},
  \bibinfo{author}{\bibfnamefont{A.}~\bibnamefont{Fujimori}},
  \bibinfo{author}{\bibfnamefont{H.}~\bibnamefont{Anzai}},
  \bibinfo{author}{\bibfnamefont{T.}~\bibnamefont{Fujita}},
  \bibinfo{author}{\bibfnamefont{A.}~\bibnamefont{Ino}},
  \bibinfo{author}{\bibfnamefont{M.}~\bibnamefont{Arita}},
  \bibinfo{author}{\bibfnamefont{H.}~\bibnamefont{Namatame}},
  \bibinfo{author}{\bibfnamefont{M.}~\bibnamefont{Taniguchi}},
  \bibinfo{author}{\bibfnamefont{Z.-X.} \bibnamefont{Shen}},
  \bibinfo{author}{\bibfnamefont{K.}~\bibnamefont{Takashima}},
  \bibinfo{author}{\bibfnamefont{K.}~\bibnamefont{Kojima}}, \bibnamefont{and}
  \bibinfo{author}{\bibfnamefont{S.-I.} \bibnamefont{Uchida}},
  \bibinfo{journal}{Phys. Rev. B} \textbf{\bibinfo{volume}{85}},
  \bibinfo{pages}{104515} (\bibinfo{year}{2012}).

\bibitem[{\citenamefont{Hoffman et~al.}(2002)\citenamefont{Hoffman, Hudson,
  Lang, Madhavan, Eisaki, Uchida, and Davis}}]{Hoffman:2002}
\bibinfo{author}{\bibfnamefont{J.~E.} \bibnamefont{Hoffman}},
  \bibinfo{author}{\bibfnamefont{E.~W.} \bibnamefont{Hudson}},
  \bibinfo{author}{\bibfnamefont{K.~M.} \bibnamefont{Lang}},
  \bibinfo{author}{\bibfnamefont{V.}~\bibnamefont{Madhavan}},
  \bibinfo{author}{\bibfnamefont{H.}~\bibnamefont{Eisaki}},
  \bibinfo{author}{\bibfnamefont{S.}~\bibnamefont{Uchida}}, \bibnamefont{and}
  \bibinfo{author}{\bibfnamefont{J.~C.} \bibnamefont{Davis}},
  \bibinfo{journal}{Nature} \textbf{\bibinfo{volume}{295}},
  \bibinfo{pages}{466} (\bibinfo{year}{2002}).

\bibitem[{\citenamefont{Howald et~al.}(2003)\citenamefont{Howald, Eisaki,
  Kaneko, Greven, and Kapitulnik}}]{Howald:2003}
\bibinfo{author}{\bibfnamefont{C.}~\bibnamefont{Howald}},
  \bibinfo{author}{\bibfnamefont{H.}~\bibnamefont{Eisaki}},
  \bibinfo{author}{\bibfnamefont{N.}~\bibnamefont{Kaneko}},
  \bibinfo{author}{\bibfnamefont{M.}~\bibnamefont{Greven}}, \bibnamefont{and}
  \bibinfo{author}{\bibfnamefont{A.}~\bibnamefont{Kapitulnik}},
  \bibinfo{journal}{Phys. Rev. B} \textbf{\bibinfo{volume}{67}},
  \bibinfo{pages}{014533} (\bibinfo{year}{2003}).

\bibitem[{\citenamefont{Wise et~al.}(2008)\citenamefont{Wise, Boyer,
  Chatterjee, Kondo, Takeuchi, Ikuta, Wang, and Hudson}}]{Wise:2008}
\bibinfo{author}{\bibfnamefont{W.~D.} \bibnamefont{Wise}},
  \bibinfo{author}{\bibfnamefont{M.~C.} \bibnamefont{Boyer}},
  \bibinfo{author}{\bibfnamefont{K.}~\bibnamefont{Chatterjee}},
  \bibinfo{author}{\bibfnamefont{T.}~\bibnamefont{Kondo}},
  \bibinfo{author}{\bibfnamefont{T.}~\bibnamefont{Takeuchi}},
  \bibinfo{author}{\bibfnamefont{H.}~\bibnamefont{Ikuta}},
  \bibinfo{author}{\bibfnamefont{Y.}~\bibnamefont{Wang}}, \bibnamefont{and}
  \bibinfo{author}{\bibfnamefont{E.~W.} \bibnamefont{Hudson}},
  \bibinfo{journal}{Nat. Phys.} \textbf{\bibinfo{volume}{4}},
  \bibinfo{pages}{696} (\bibinfo{year}{2008}).

\bibitem[{\citenamefont{Parker et~al.}(2010)\citenamefont{Parker, Aynajian,
  da~Silva~Neto, Pushp, Ono, Wen, Xu, Gu, and Yazdani}}]{Parker:2010nat}
\bibinfo{author}{\bibfnamefont{C.~V.} \bibnamefont{Parker}},
  \bibinfo{author}{\bibfnamefont{P.}~\bibnamefont{Aynajian}},
  \bibinfo{author}{\bibfnamefont{E.~H.} \bibnamefont{da~Silva~Neto}},
  \bibinfo{author}{\bibfnamefont{A.}~\bibnamefont{Pushp}},
  \bibinfo{author}{\bibfnamefont{S.}~\bibnamefont{Ono}},
  \bibinfo{author}{\bibfnamefont{J.}~\bibnamefont{Wen}},
  \bibinfo{author}{\bibfnamefont{Z.}~\bibnamefont{Xu}},
  \bibinfo{author}{\bibfnamefont{G.}~\bibnamefont{Gu}}, \bibnamefont{and}
  \bibinfo{author}{\bibfnamefont{A.}~\bibnamefont{Yazdani}},
  \bibinfo{journal}{Nature} \textbf{\bibinfo{volume}{468}},
  \bibinfo{pages}{677} (\bibinfo{year}{2010}).

\bibitem[{\citenamefont{Lawler et~al.}(2010)\citenamefont{Lawler, Fujita, Lee,
  Schmidt, Kohsaka, Kim, Eisaki, Uchida, Davis, Sethna, and
  Kim}}]{Lawler:2010n}
\bibinfo{author}{\bibfnamefont{M.~J.} \bibnamefont{Lawler}},
  \bibinfo{author}{\bibfnamefont{K.}~\bibnamefont{Fujita}},
  \bibinfo{author}{\bibfnamefont{J.}~\bibnamefont{Lee}},
  \bibinfo{author}{\bibfnamefont{A.~R.} \bibnamefont{Schmidt}},
  \bibinfo{author}{\bibfnamefont{Y.}~\bibnamefont{Kohsaka}},
  \bibinfo{author}{\bibfnamefont{C.~K.} \bibnamefont{Kim}},
  \bibinfo{author}{\bibfnamefont{H.}~\bibnamefont{Eisaki}},
  \bibinfo{author}{\bibfnamefont{S.}~\bibnamefont{Uchida}},
  \bibinfo{author}{\bibfnamefont{J.~C.} \bibnamefont{Davis}},
  \bibinfo{author}{\bibfnamefont{J.~P.} \bibnamefont{Sethna}},
  \bibnamefont{and} \bibinfo{author}{\bibfnamefont{E.-A.} \bibnamefont{Kim}},
  \bibinfo{journal}{Nature} \textbf{\bibinfo{volume}{466}},
  \bibinfo{pages}{347} (\bibinfo{year}{2010}).

\bibitem[{\citenamefont{Mesaros et~al.}(2011)\citenamefont{Mesaros, Fujita,
  Eisaki, Uchida, Davis, Sachdev, Zaanen, Lawler, and Kim}}]{Mesaros:2011s}
\bibinfo{author}{\bibfnamefont{A.}~\bibnamefont{Mesaros}},
  \bibinfo{author}{\bibfnamefont{K.}~\bibnamefont{Fujita}},
  \bibinfo{author}{\bibfnamefont{H.}~\bibnamefont{Eisaki}},
  \bibinfo{author}{\bibfnamefont{S.}~\bibnamefont{Uchida}},
  \bibinfo{author}{\bibfnamefont{J.~C.} \bibnamefont{Davis}},
  \bibinfo{author}{\bibfnamefont{S.}~\bibnamefont{Sachdev}},
  \bibinfo{author}{\bibfnamefont{J.}~\bibnamefont{Zaanen}},
  \bibinfo{author}{\bibfnamefont{M.~J.} \bibnamefont{Lawler}},
  \bibnamefont{and} \bibinfo{author}{\bibfnamefont{E.-A.} \bibnamefont{Kim}},
  \bibinfo{journal}{Science} \textbf{\bibinfo{volume}{333}},
  \bibinfo{pages}{426} (\bibinfo{year}{2011}).

\bibitem[{\citenamefont{da~Silva~Neto et~al.}(2012)\citenamefont{da~Silva~Neto,
  Parker, Aynajian, Pushp, Yazdani, Wen, Xu, and Gu}}]{Eduardo:2012prb}
\bibinfo{author}{\bibfnamefont{E.~H.} \bibnamefont{da~Silva~Neto}},
  \bibinfo{author}{\bibfnamefont{C.~V.} \bibnamefont{Parker}},
  \bibinfo{author}{\bibfnamefont{P.}~\bibnamefont{Aynajian}},
  \bibinfo{author}{\bibfnamefont{A.}~\bibnamefont{Pushp}},
  \bibinfo{author}{\bibfnamefont{A.}~\bibnamefont{Yazdani}},
  \bibinfo{author}{\bibfnamefont{J.}~\bibnamefont{Wen}},
  \bibinfo{author}{\bibfnamefont{Z.}~\bibnamefont{Xu}}, \bibnamefont{and}
  \bibinfo{author}{\bibfnamefont{G.}~\bibnamefont{Gu}}, \bibinfo{journal}{Phys.
  Rev. B} \textbf{\bibinfo{volume}{85}}, \bibinfo{pages}{104521}
  (\bibinfo{year}{2012}).

\bibitem[{\citenamefont{Emery}(1987)}]{Emery:1987prl}
\bibinfo{author}{\bibfnamefont{V.~J.} \bibnamefont{Emery}},
  \bibinfo{journal}{Phys. Rev. Lett.} \textbf{\bibinfo{volume}{58}},
  \bibinfo{pages}{2794} (\bibinfo{year}{1987}).

\bibitem[{\citenamefont{Littlewood et~al.}(1989)\citenamefont{Littlewood,
  Varma, Schmitt-Rink, and Abrahams}}]{Littlewood:1989}
\bibinfo{author}{\bibfnamefont{P.~B.} \bibnamefont{Littlewood}},
  \bibinfo{author}{\bibfnamefont{C.~M.} \bibnamefont{Varma}},
  \bibinfo{author}{\bibfnamefont{S.}~\bibnamefont{Schmitt-Rink}},
  \bibnamefont{and} \bibinfo{author}{\bibfnamefont{E.}~\bibnamefont{Abrahams}},
  \bibinfo{journal}{Phys. Rev. B} \textbf{\bibinfo{volume}{39}},
  \bibinfo{pages}{12371} (\bibinfo{year}{1989}).

\bibitem[{\citenamefont{Littlewood}(1990)}]{Littlewood:1990}
\bibinfo{author}{\bibfnamefont{P.~B.} \bibnamefont{Littlewood}},
  \bibinfo{journal}{Phys. Rev. B} \textbf{\bibinfo{volume}{42}},
  \bibinfo{pages}{10075} (\bibinfo{year}{1990}).

\bibitem[{\citenamefont{Fischer and Kim}(2011)}]{Fischer:2011}
\bibinfo{author}{\bibfnamefont{M.~H.} \bibnamefont{Fischer}} \bibnamefont{and}
  \bibinfo{author}{\bibfnamefont{E.-A.} \bibnamefont{Kim}},
  \bibinfo{journal}{Phys. Rev. B} \textbf{\bibinfo{volume}{84}},
  \bibinfo{pages}{144502} (\bibinfo{year}{2011}).

\bibitem[{\citenamefont{Metlitski and Sachdev}(2010)}]{Metlitski:2010vf}
\bibinfo{author}{\bibfnamefont{M.~A.} \bibnamefont{Metlitski}}
  \bibnamefont{and} \bibinfo{author}{\bibfnamefont{S.}~\bibnamefont{Sachdev}},
  \bibinfo{journal}{New J. Phys.} \textbf{\bibinfo{volume}{12}},
  \bibinfo{pages}{105007} (\bibinfo{year}{2010}).

\bibitem[{\citenamefont{Melikyan and Norman}(2011)}]{Melikyan:2011}
\bibinfo{author}{\bibfnamefont{A.}~\bibnamefont{Melikyan}} \bibnamefont{and}
  \bibinfo{author}{\bibfnamefont{M.~R.} \bibnamefont{Norman}}
  (\bibinfo{year}{2011}), \eprint{http://arxiv.org/abs/1102.5443}.

\bibitem[{\citenamefont{Husemann and Metzner}(2012)}]{Husemann:2012vg}
\bibinfo{author}{\bibfnamefont{C.}~\bibnamefont{Husemann}} \bibnamefont{and}
  \bibinfo{author}{\bibfnamefont{W.}~\bibnamefont{Metzner}},
  \bibinfo{journal}{Phys. Rev. B} \textbf{\bibinfo{volume}{86}},
  \bibinfo{pages}{085113} (\bibinfo{year}{2012}).

\bibitem[{\citenamefont{Holder and Metzner}(2012)}]{Holder:2012ks}
\bibinfo{author}{\bibfnamefont{T.}~\bibnamefont{Holder}} \bibnamefont{and}
  \bibinfo{author}{\bibfnamefont{W.}~\bibnamefont{Metzner}},
  \bibinfo{journal}{Phys. Rev. B} \textbf{\bibinfo{volume}{85}},
  \bibinfo{pages}{165130} (\bibinfo{year}{2012}).

\bibitem[{\citenamefont{Efetov et~al.}(2013)\citenamefont{Efetov, Meier, and
  P\'{e}pin}}]{Efetov:2013}
\bibinfo{author}{\bibfnamefont{K.~B.} \bibnamefont{Efetov}},
  \bibinfo{author}{\bibfnamefont{H.}~\bibnamefont{Meier}}, \bibnamefont{and}
  \bibinfo{author}{\bibfnamefont{C.}~\bibnamefont{P\'{e}pin}},
  \bibinfo{journal}{Nature Physics} \textbf{\bibinfo{volume}{9}},
  \bibinfo{pages}{442} (\bibinfo{year}{2013}).

\bibitem[{\citenamefont{Sachdev and La~Placa}(2013)}]{Sachdev:2013}
\bibinfo{author}{\bibfnamefont{S.}~\bibnamefont{Sachdev}} \bibnamefont{and}
  \bibinfo{author}{\bibfnamefont{R.}~\bibnamefont{La~Placa}},
  \bibinfo{journal}{Phys. Rev. Lett.} \textbf{\bibinfo{volume}{111}},
  \bibinfo{pages}{027202} (\bibinfo{year}{2013}).

\bibitem[{\citenamefont{Meier et~al.}(2013)\citenamefont{Meier, Einenkel,
  P\'epin, and Efetov}}]{Meier:2013}
\bibinfo{author}{\bibfnamefont{H.}~\bibnamefont{Meier}},
  \bibinfo{author}{\bibfnamefont{M.}~\bibnamefont{Einenkel}},
  \bibinfo{author}{\bibfnamefont{C.}~\bibnamefont{P\'epin}}, \bibnamefont{and}
  \bibinfo{author}{\bibfnamefont{K.~B.} \bibnamefont{Efetov}},
  \bibinfo{journal}{Phys. Rev. B} \textbf{\bibinfo{volume}{88}},
  \bibinfo{pages}{020506} (\bibinfo{year}{2013}).

\bibitem[{\citenamefont{Scalettar et~al.}(1991)\citenamefont{Scalettar,
  Scalapino, Sugar, and White}}]{Scalettar:1991}
\bibinfo{author}{\bibfnamefont{R.~T.} \bibnamefont{Scalettar}},
  \bibinfo{author}{\bibfnamefont{D.~J.} \bibnamefont{Scalapino}},
  \bibinfo{author}{\bibfnamefont{R.~L.} \bibnamefont{Sugar}}, \bibnamefont{and}
  \bibinfo{author}{\bibfnamefont{S.~R.} \bibnamefont{White}},
  \bibinfo{journal}{Phys. Rev. B} \textbf{\bibinfo{volume}{44}},
  \bibinfo{pages}{770} (\bibinfo{year}{1991}).

\bibitem[{\citenamefont{Kivelson et~al.}(2004)\citenamefont{Kivelson, Fradkin,
  and Geballe}}]{Kivelson:2004}
\bibinfo{author}{\bibfnamefont{S.~A.} \bibnamefont{Kivelson}},
  \bibinfo{author}{\bibfnamefont{E.}~\bibnamefont{Fradkin}}, \bibnamefont{and}
  \bibinfo{author}{\bibfnamefont{T.~H.} \bibnamefont{Geballe}},
  \bibinfo{journal}{Phys. Rev. B} \textbf{\bibinfo{volume}{69}},
  \bibinfo{pages}{144505} (\bibinfo{year}{2004}).

\bibitem[{\citenamefont{Sun et~al.}(2008)\citenamefont{Sun, Fregoso, Lawler,
  and Fradkin}}]{Sun:2008}
\bibinfo{author}{\bibfnamefont{K.}~\bibnamefont{Sun}},
  \bibinfo{author}{\bibfnamefont{B.~M.} \bibnamefont{Fregoso}},
  \bibinfo{author}{\bibfnamefont{M.~J.} \bibnamefont{Lawler}},
  \bibnamefont{and} \bibinfo{author}{\bibfnamefont{E.}~\bibnamefont{Fradkin}},
  \bibinfo{journal}{Phys. Rev. B} \textbf{\bibinfo{volume}{78}},
  \bibinfo{pages}{085124} (\bibinfo{year}{2008}).

\bibitem[{\citenamefont{Andersen et~al.}(1995)\citenamefont{Andersen,
  Liechtenstein, Jepsen, and Paulsen}}]{Andersen:1995}
\bibinfo{author}{\bibfnamefont{O.}~\bibnamefont{Andersen}},
  \bibinfo{author}{\bibfnamefont{A.}~\bibnamefont{Liechtenstein}},
  \bibinfo{author}{\bibfnamefont{O.}~\bibnamefont{Jepsen}}, \bibnamefont{and}
  \bibinfo{author}{\bibfnamefont{F.}~\bibnamefont{Paulsen}},
  \bibinfo{journal}{J. Phys. Chem. Solids} \textbf{\bibinfo{volume}{56}},
  \bibinfo{pages}{1573 } (\bibinfo{year}{1995}).

\bibitem[{\citenamefont{Arrigoni et~al.}(2009)\citenamefont{Arrigoni, Aichhorn,
  Daghofer, and Hanke}}]{Arrigoni:2009njp}
\bibinfo{author}{\bibfnamefont{E.}~\bibnamefont{Arrigoni}},
  \bibinfo{author}{\bibfnamefont{M.}~\bibnamefont{Aichhorn}},
  \bibinfo{author}{\bibfnamefont{M.}~\bibnamefont{Daghofer}}, \bibnamefont{and}
  \bibinfo{author}{\bibfnamefont{W.}~\bibnamefont{Hanke}},
  \bibinfo{journal}{New J. Phys.} \textbf{\bibinfo{volume}{11}},
  \bibinfo{pages}{055066} (\bibinfo{year}{2009}).

\bibitem[{\citenamefont{Hybertsen et~al.}(1989)\citenamefont{Hybertsen,
  Schl\"uter, and Christensen}}]{Hybertsen:1989prb}
\bibinfo{author}{\bibfnamefont{M.~S.} \bibnamefont{Hybertsen}},
  \bibinfo{author}{\bibfnamefont{M.}~\bibnamefont{Schl\"uter}},
  \bibnamefont{and} \bibinfo{author}{\bibfnamefont{N.~E.}
  \bibnamefont{Christensen}}, \bibinfo{journal}{Phys. Rev. B}
  \textbf{\bibinfo{volume}{39}}, \bibinfo{pages}{9028} (\bibinfo{year}{1989}).

\bibitem[{\citenamefont{Carrington and Yelland}(2007)}]{Carrington:2007}
\bibinfo{author}{\bibfnamefont{A.}~\bibnamefont{Carrington}} \bibnamefont{and}
  \bibinfo{author}{\bibfnamefont{E.~A.} \bibnamefont{Yelland}},
  \bibinfo{journal}{Phys. Rev. B} \textbf{\bibinfo{volume}{76}},
  \bibinfo{pages}{140508} (\bibinfo{year}{2007}).

\bibitem[{\citenamefont{Bang et~al.}(1993)\citenamefont{Bang, Kotliar,
  Raimondi, Castellani, and Grilli}}]{Bang:1993}
\bibinfo{author}{\bibfnamefont{Y.}~\bibnamefont{Bang}},
  \bibinfo{author}{\bibfnamefont{G.}~\bibnamefont{Kotliar}},
  \bibinfo{author}{\bibfnamefont{R.}~\bibnamefont{Raimondi}},
  \bibinfo{author}{\bibfnamefont{C.}~\bibnamefont{Castellani}},
  \bibnamefont{and} \bibinfo{author}{\bibfnamefont{M.}~\bibnamefont{Grilli}},
  \bibinfo{journal}{Phys. Rev. B} \textbf{\bibinfo{volume}{47}},
  \bibinfo{pages}{3323} (\bibinfo{year}{1993}).

\bibitem[{\citenamefont{Yamase}(2004)}]{Yamase:2004}
\bibinfo{author}{\bibfnamefont{H.}~\bibnamefont{Yamase}},
  \bibinfo{journal}{Phys. Rev. Lett.} \textbf{\bibinfo{volume}{93}},
  \bibinfo{pages}{266404} (\bibinfo{year}{2004}).

\bibitem[{\citenamefont{Yamase et~al.}(2005)\citenamefont{Yamase, Oganesyan,
  and Metzner}}]{Yamase:2005}
\bibinfo{author}{\bibfnamefont{H.}~\bibnamefont{Yamase}},
  \bibinfo{author}{\bibfnamefont{V.}~\bibnamefont{Oganesyan}},
  \bibnamefont{and} \bibinfo{author}{\bibfnamefont{W.}~\bibnamefont{Metzner}},
  \bibinfo{journal}{Phys. Rev. B} \textbf{\bibinfo{volume}{72}},
  \bibinfo{pages}{035114} (\bibinfo{year}{2005}).

\bibitem[{\citenamefont{Damascelli et~al.}(2003)\citenamefont{Damascelli,
  Hussain, and Shen}}]{Damascelli:2003}
\bibinfo{author}{\bibfnamefont{A.}~\bibnamefont{Damascelli}},
  \bibinfo{author}{\bibfnamefont{Z.}~\bibnamefont{Hussain}}, \bibnamefont{and}
  \bibinfo{author}{\bibfnamefont{Z.-X.} \bibnamefont{Shen}},
  \bibinfo{journal}{Rev. Mod. Phys.} \textbf{\bibinfo{volume}{75}},
  \bibinfo{pages}{473} (\bibinfo{year}{2003}).

\bibitem[{\citenamefont{Atkinson et~al.}(2012)\citenamefont{Atkinson, Bazak,
  and Andersen}}]{Atkinson:2012}
\bibinfo{author}{\bibfnamefont{W.~A.} \bibnamefont{Atkinson}},
  \bibinfo{author}{\bibfnamefont{J.~D.} \bibnamefont{Bazak}}, \bibnamefont{and}
  \bibinfo{author}{\bibfnamefont{B.~M.} \bibnamefont{Andersen}},
  \bibinfo{journal}{Phys. Rev. Lett.} \textbf{\bibinfo{volume}{109}},
  \bibinfo{pages}{267004} (\bibinfo{year}{2012}).

\bibitem[{\citenamefont{Hayden}()}]{Hayden:2013MM}
\bibinfo{author}{\bibfnamefont{S.}~\bibnamefont{Hayden}},
  \bibinfo{howpublished}{APS March Meeting},
  \urlprefix\url{http://meetings.aps.org/link/BAPS.2013.MAR.B2.2}.

\bibitem[{\citenamefont{Kohsaka et~al.}(2008)\citenamefont{Kohsaka, Taylor,
  Wahl, Schmidt, Lee, Fujita, Alldredge, McElroy, Lee, Eisaki, Uchida, Lee, and
  Davis}}]{Kohsaka:2008}
\bibinfo{author}{\bibfnamefont{Y.}~\bibnamefont{Kohsaka}},
  \bibinfo{author}{\bibfnamefont{C.}~\bibnamefont{Taylor}},
  \bibinfo{author}{\bibfnamefont{P.}~\bibnamefont{Wahl}},
  \bibinfo{author}{\bibfnamefont{A.}~\bibnamefont{Schmidt}},
  \bibinfo{author}{\bibfnamefont{J.}~\bibnamefont{Lee}},
  \bibinfo{author}{\bibfnamefont{K.}~\bibnamefont{Fujita}},
  \bibinfo{author}{\bibfnamefont{J.~W.} \bibnamefont{Alldredge}},
  \bibinfo{author}{\bibfnamefont{K.}~\bibnamefont{McElroy}},
  \bibinfo{author}{\bibfnamefont{J.}~\bibnamefont{Lee}},
  \bibinfo{author}{\bibfnamefont{H.}~\bibnamefont{Eisaki}},
  \bibinfo{author}{\bibfnamefont{S.}~\bibnamefont{Uchida}},
  \bibinfo{author}{\bibfnamefont{D.-H.} \bibnamefont{Lee}}, \bibnamefont{and}
  \bibinfo{author}{\bibfnamefont{J.~C.} \bibnamefont{Davis}},
  \bibinfo{journal}{Nature} \textbf{\bibinfo{volume}{454}},
  \bibinfo{pages}{1072} (\bibinfo{year}{2008}).

\bibitem[{\citenamefont{Hanaguri et~al.}(2004)\citenamefont{Hanaguri, Lupien,
  Kohsaka, Lee, Azuma, Takano, Takagi, and Davis}}]{Hanaguri:2004}
\bibinfo{author}{\bibfnamefont{T.}~\bibnamefont{Hanaguri}},
  \bibinfo{author}{\bibfnamefont{C.}~\bibnamefont{Lupien}},
  \bibinfo{author}{\bibfnamefont{Y.}~\bibnamefont{Kohsaka}},
  \bibinfo{author}{\bibfnamefont{D.}~\bibnamefont{Lee}},
  \bibinfo{author}{\bibfnamefont{M.}~\bibnamefont{Azuma}},
  \bibinfo{author}{\bibfnamefont{M.}~\bibnamefont{Takano}},
  \bibinfo{author}{\bibfnamefont{H.}~\bibnamefont{Takagi}}, \bibnamefont{and}
  \bibinfo{author}{\bibfnamefont{J.}~\bibnamefont{Davis}},
  \bibinfo{journal}{Nature} \textbf{\bibinfo{volume}{430}},
  \bibinfo{pages}{1001} (\bibinfo{year}{2004}).

\bibitem[{\citenamefont{Riggs et~al.}(2011)\citenamefont{Riggs, Vafek, Kemper,
  Betts, Migliori, Balakirev, Hardy, Liang, Bonn, and
  Boebinger}}]{Riggs:2011ii}
\bibinfo{author}{\bibfnamefont{S.~C.} \bibnamefont{Riggs}},
  \bibinfo{author}{\bibfnamefont{O.}~\bibnamefont{Vafek}},
  \bibinfo{author}{\bibfnamefont{J.~B.} \bibnamefont{Kemper}},
  \bibinfo{author}{\bibfnamefont{J.~B.} \bibnamefont{Betts}},
  \bibinfo{author}{\bibfnamefont{A.}~\bibnamefont{Migliori}},
  \bibinfo{author}{\bibfnamefont{F.~F.} \bibnamefont{Balakirev}},
  \bibinfo{author}{\bibfnamefont{W.~N.} \bibnamefont{Hardy}},
  \bibinfo{author}{\bibfnamefont{R.}~\bibnamefont{Liang}},
  \bibinfo{author}{\bibfnamefont{D.~A.} \bibnamefont{Bonn}}, \bibnamefont{and}
  \bibinfo{author}{\bibfnamefont{G.~S.} \bibnamefont{Boebinger}},
  \bibinfo{journal}{Nat. Phys.} \textbf{\bibinfo{volume}{7}},
  \bibinfo{pages}{332} (\bibinfo{year}{2011}).

\bibitem[{\citenamefont{Singleton et~al.}(2010)\citenamefont{Singleton, de~la
  Cruz, McDonald, Li, Altarawneh, Goddard, Franke, Rickel, Mielke, Yao, and
  Dai}}]{Singleton:2010bz}
\bibinfo{author}{\bibfnamefont{J.}~\bibnamefont{Singleton}},
  \bibinfo{author}{\bibfnamefont{C.}~\bibnamefont{de~la Cruz}},
  \bibinfo{author}{\bibfnamefont{R.~D.} \bibnamefont{McDonald}},
  \bibinfo{author}{\bibfnamefont{S.}~\bibnamefont{Li}},
  \bibinfo{author}{\bibfnamefont{M.}~\bibnamefont{Altarawneh}},
  \bibinfo{author}{\bibfnamefont{P.}~\bibnamefont{Goddard}},
  \bibinfo{author}{\bibfnamefont{I.}~\bibnamefont{Franke}},
  \bibinfo{author}{\bibfnamefont{D.}~\bibnamefont{Rickel}},
  \bibinfo{author}{\bibfnamefont{C.~H.} \bibnamefont{Mielke}},
  \bibinfo{author}{\bibfnamefont{X.}~\bibnamefont{Yao}}, \bibnamefont{and}
  \bibinfo{author}{\bibfnamefont{P.}~\bibnamefont{Dai}},
  \bibinfo{journal}{Phys. Rev. Lett.} \textbf{\bibinfo{volume}{104}},
  \bibinfo{pages}{086403} (\bibinfo{year}{2010}).

\bibitem[{\citenamefont{Bangura et~al.}(2008)\citenamefont{Bangura, Fletcher,
  Carrington, Levallois, Nardone, Vignolle, Heard, Doiron-Leyraud, LeBoeuf,
  Taillefer, Adachi, Proust, and Hussey}}]{Bangura:2008cg}
\bibinfo{author}{\bibfnamefont{A.~F.} \bibnamefont{Bangura}},
  \bibinfo{author}{\bibfnamefont{J.~D.} \bibnamefont{Fletcher}},
  \bibinfo{author}{\bibfnamefont{A.}~\bibnamefont{Carrington}},
  \bibinfo{author}{\bibfnamefont{J.}~\bibnamefont{Levallois}},
  \bibinfo{author}{\bibfnamefont{M.}~\bibnamefont{Nardone}},
  \bibinfo{author}{\bibfnamefont{B.}~\bibnamefont{Vignolle}},
  \bibinfo{author}{\bibfnamefont{P.~J.} \bibnamefont{Heard}},
  \bibinfo{author}{\bibfnamefont{N.}~\bibnamefont{Doiron-Leyraud}},
  \bibinfo{author}{\bibfnamefont{D.}~\bibnamefont{LeBoeuf}},
  \bibinfo{author}{\bibfnamefont{L.}~\bibnamefont{Taillefer}},
  \bibinfo{author}{\bibfnamefont{S.}~\bibnamefont{Adachi}},
  \bibinfo{author}{\bibfnamefont{C.}~\bibnamefont{Proust}}, \bibnamefont{and}
  \bibinfo{author}{\bibfnamefont{N.~E.} \bibnamefont{Hussey}},
  \bibinfo{journal}{Phys. Rev. Lett.} \textbf{\bibinfo{volume}{100}},
  \bibinfo{pages}{047004} (\bibinfo{year}{2008}).

\bibitem[{\citenamefont{Yelland et~al.}(2008)\citenamefont{Yelland, Singleton,
  Mielke, Harrison, Balakirev, Dabrowski, and Cooper}}]{Yelland:2008dh}
\bibinfo{author}{\bibfnamefont{E.}~\bibnamefont{Yelland}},
  \bibinfo{author}{\bibfnamefont{J.}~\bibnamefont{Singleton}},
  \bibinfo{author}{\bibfnamefont{C.}~\bibnamefont{Mielke}},
  \bibinfo{author}{\bibfnamefont{N.}~\bibnamefont{Harrison}},
  \bibinfo{author}{\bibfnamefont{F.}~\bibnamefont{Balakirev}},
  \bibinfo{author}{\bibfnamefont{B.}~\bibnamefont{Dabrowski}},
  \bibnamefont{and} \bibinfo{author}{\bibfnamefont{J.}~\bibnamefont{Cooper}},
  \bibinfo{journal}{Phys. Rev. Lett.} \textbf{\bibinfo{volume}{100}},
  \bibinfo{pages}{047003} (\bibinfo{year}{2008}).

\bibitem[{\citenamefont{Sebastian et~al.}(2008)\citenamefont{Sebastian,
  Harrison, Palm, Murphy, Mielke, Liang, Bonn, Hardy, and
  Lonzarich}}]{Sebastian:2008hp}
\bibinfo{author}{\bibfnamefont{S.~E.} \bibnamefont{Sebastian}},
  \bibinfo{author}{\bibfnamefont{N.}~\bibnamefont{Harrison}},
  \bibinfo{author}{\bibfnamefont{E.}~\bibnamefont{Palm}},
  \bibinfo{author}{\bibfnamefont{T.~P.} \bibnamefont{Murphy}},
  \bibinfo{author}{\bibfnamefont{C.~H.} \bibnamefont{Mielke}},
  \bibinfo{author}{\bibfnamefont{R.}~\bibnamefont{Liang}},
  \bibinfo{author}{\bibfnamefont{D.~A.} \bibnamefont{Bonn}},
  \bibinfo{author}{\bibfnamefont{W.~N.} \bibnamefont{Hardy}}, \bibnamefont{and}
  \bibinfo{author}{\bibfnamefont{G.~G.} \bibnamefont{Lonzarich}},
  \bibinfo{journal}{Nature} \textbf{\bibinfo{volume}{454}},
  \bibinfo{pages}{200} (\bibinfo{year}{2008}).

\bibitem[{\citenamefont{Sebastian
  et~al.}(2012{\natexlab{b}})\citenamefont{Sebastian, Harrison, Liang, Bonn,
  Hardy, Mielke, and Lonzarich}}]{Sebastian:2012bl}
\bibinfo{author}{\bibfnamefont{S.}~\bibnamefont{Sebastian}},
  \bibinfo{author}{\bibfnamefont{N.}~\bibnamefont{Harrison}},
  \bibinfo{author}{\bibfnamefont{R.}~\bibnamefont{Liang}},
  \bibinfo{author}{\bibfnamefont{D.}~\bibnamefont{Bonn}},
  \bibinfo{author}{\bibfnamefont{W.}~\bibnamefont{Hardy}},
  \bibinfo{author}{\bibfnamefont{C.}~\bibnamefont{Mielke}}, \bibnamefont{and}
  \bibinfo{author}{\bibfnamefont{G.}~\bibnamefont{Lonzarich}},
  \bibinfo{journal}{Phys. Rev. Lett.} \textbf{\bibinfo{volume}{108}},
  \bibinfo{pages}{196403} (\bibinfo{year}{2012}{\natexlab{b}}).

\end{thebibliography}

\end{document}